\documentclass[10pt]{article}

\usepackage[latin1]{inputenc}
\usepackage{epsfig}
\usepackage{a4wide}
\usepackage{latexsym}
\usepackage{amssymb}
\usepackage{amsmath}
\usepackage{graphics}

\begin{document}

\def\O{{\cal O}}
\def\N{{\cal N}}
\def\enu{\epsilon_\nu}
\def\hxi{\hat x_i}
\def\hatx{{\bf \hat x}}
\def\d{{\rm d}}
\def\i{{\rm i}}
\def\e{{\bf e}}
\def\ex{{\rm e}}
\def\x{{\bf x}}
\def\X{{\bf X}}
\def\J{{\rm J}}
\def\r{{\bf r}}
\def\s{{\bf s}}
\def\k{{\bf k}}
\def\y{{\bf y}}
\def\p{{\bf p}}
\def\q{{\bf q}}
\def\z{{\bf z}}
\def\R{{\bf R}}
\def\A{{\bf A}}
\def\v{{\bf v}}
\def\u{{\bf u}}
\def\w{{\bf w}}
\def\U{{\bf U}}
\def\cm{{\rm cm}}
\def\l{{\bf l}}
\def\sec{{\rm sec}}
\def\Ckol{C_{Kol}}
\def\flux{\bar\epsilon}
\def\b{b_{kpq}}
\def\smalsev{{\scriptscriptstyle \frac{7}{3} }}
\def\smalfi{{\scriptscriptstyle \frac{5}{3} }}
\def\smalL{{\scriptscriptstyle{\rm L}}}
\def\smalP{{\scriptscriptstyle {\rm P}}}
\def\smalT{{\scriptscriptstyle {\rm T}}}
\def\smalE{{\scriptscriptstyle{\rm E}}}
\def\smal1n{{\scriptscriptstyle (1,n)}}
\def\smaln{{\scriptscriptstyle (n)}}
\def\smalA{{\scriptscriptstyle {\rm A}}}
\def\smalze{{\scriptscriptstyle (0)}}
\def\smalun{{\scriptscriptstyle (1)}}
\def\smaldu{{\scriptscriptstyle (2)}}
\def\smaln{{\scriptscriptstyle (n)}}
\def\UP{U^\smalP}
\def\gammaP{\gamma^\smalP}
\def\shell{{\tt S}}
\def\ball{{\tt B}}
\def\nav{\bar N}
\def\micron{\mu{\rm m}}
\font\brm=cmr10 at 24truept
\font\bfm=cmbx10 at 15truept

\baselineskip 0.7cm

\centerline{\brm Particle transport in a random velocity}
\centerline{\brm field with Lagrangian statistics} 
\vskip 20pt
\centerline{Piero Olla}
\vskip 5pt
\centerline{ISAC-CNR}
\centerline{Sezione di Lecce}
\centerline{73100 Lecce Italy}
\vskip 20pt

\centerline{\bf Abstract}
\vskip 5pt
The transport properties of a random velocity field with Kolmogorov 
spectrum and time correlations defined along Lagrangian trajectories are analyzed. 
The analysis is carried on in the limit of short correlation times, as a perturbation 
theory in the ratio, scale by scale, of the eddy decay and turn-over time.
Various quantities such as the Batchelor constant and the dimensionless constants
entering the expression for particle relative and self-diffusion are given in terms 
of this ratio and of the Kolmogorov constant.
Particular attention is paid to particles with finite inertia. The
self-diffusion properties of a particle with Stokes time longer than the Kolmogorov 
time are determined, verifying on an analytical example the dimensional results
of $[$nlin.CD/0103018$]$. Expressions for the fluid velocity Lagrangian correlations
and correlation times along a solid particle trajectory, are provided in several 
parameter regimes, including the infinite Stokes time limit corresponding to 
Eulerian correlations. The concentration fluctuation spectrum and the 
non-ergodic properties of a suspension of heavy particles in a turbulent flow, 
in the same regime, are analyzed. The concentration spectrum is predicted to obey,
above the scale of eddies with lifetime equal to the Stokes time, 
a power law with universal $-4/3$ exponent, 
and to be otherwise independent of the nature of the turbulent flow.
A preference of the solid particle to lie in less energetic regions of the flow is observed.

\vfill\eject

\centerline{\bf I. Introduction}
\vskip 5pt
One of the differences between high Reynolds number turbulence and other examples of random fields 
with power law scaling,
is the Lagrangian nature of time 
correlations \cite{kraichnan64}. 
From the theoretical point of view, the need for a Lagrangian treatment of time correlations has 
been one of the main difficulties in the realization of statistical turbulent closures 
\cite{kraichnan65}. Because of
this, many such theories assume from the start that the turbulence dynamics be equivalent to that
of a random velocity field with identical energy spectrum but Eulerian time statistics, i.e. 
the fluctuations decay without being transported by the larger vortices
\cite{kraichnan71,orszag77,yakhot86}.
Such an assumption does not work in the case of particle transport: both relative and
self-diffusion are affected by the way in which time correlations are defined.

Concerning self-diffusion, in Kolmogorov turbulence, fluctuations at a scale $l$ within the 
inertial range, have characteristic velocity $\sim l^\frac{1}{3}$ and decay time 
$\sim l^{-\frac{2}{3}}$ along fluid trajectories. Hence, in a time $t$ the velocity of a fluid 
parcel will change by an amount of the order of that of a fluctuation with that lifetime, i.e. 
by $t^\frac{1}{2}$. If the fluctuations were not advected by the flow, the fluid parcel would 
see the fluctuation only for the time $\sim l^{-1}$ it takes to cross it. The variation of 
the fluid parcel velocity in a time $t$ would be therefore $\sim t^\frac{1}{3}$. 

Concerning relative diffusion, this process is determined by vortices with the size of the fluid 
parcel separation at the given time. If these vortices were fixed in space, their effect on 
relative diffusion would be proportional to the crossing time by the fluid parcels, which is 
determined by the large scale properties of the flow. In other words, if time correlations were
given in an Eulerian reference frame, the process of relative diffusion would not depend solely 
on the inter-particle distance and on the velocity difference, but also on the 
total velocity.

Given the difficulty in defining a velocity field with Lagrangian statistics, a successful 
strategy for the treatment of transport has been to neglect time correlations altogether, i.e. to consider a velocity
field $\u$ such that $\langle u_\alpha (\x,t)u_\beta (0,0)\rangle = 
U_{\alpha\beta}(\x)\delta(t)$: the
so called Kraichnan model \cite{kraichnan94}. In this model, Eulerian and
Lagrangian time statistics trivially coincide in what is the zero order of some perturbation
theory in powers of the correlation time of the turbulence. It has been possible, in particular,
to determine the anomalous scaling exponents of a passive scalar injected at large scales in 
the velocity field \cite{chertkov95,gawedzki95,frisch98,gat98}. The origin of this success is that, 
although the time structure of the
velocity correlation is lost, that of the relative displacement, whose geometrical properties
determine the passive scalar correlations, is preserved \cite{gat98a,pumir00,celani01}. 
(For instance, particle pair separation still obeys Richardson diffusion).

The question, at this point, is how to introduce finite correlation times in a perturbative
manner, but preserving the Lagrangian nature of correlations. There are practical reasons to
do this. One motivation, of course, is to be able to determine the time correlations
of the particle velocities. Lagrangian dispersion models \cite{thomson87,thomson90,borgas94} are 
based on the adoption of 
prescriptions on the form of these time correlations; to be able to determine them 
directly from the statistical properties of the velocity field would be therefore of some interest.

It must be said that most of the prescriptions entering a Lagrangian dispersion model
could be obtained, in practice, by dimensional reasoning or by experiments. 
In some cases, like in the presence of particles endowed with inertia,
this turns out, however, to be a difficult task \cite{csanady63,sawford91a}.  
It is very difficult, for instance, to
make assumptions
on the preference of solid particles to lie in certain regions of the flow instead of others
\cite{wang92,elgobashi92,wang93}.
Solid particle transport by a turbulent flow is an example of a situation in which careful
treatment of the time dependent statistics
of the velocity field is essential.
It is precisely the interplay
between the response time of the solid particle to the fluid, i.e. the Stokes time
$\tau_S$, and the characteristic times of the turbulent flow \cite{olla01}, which determines
the dynamics, and this is clearly lost when all the turbulent times are sent to zero. 

Recently, there has been strong theoretical interest on the problem of turbulence induced 
concentration fluctuations in a heavy particle suspension. In \cite{elperin00}, the role
of a finite correlation time of the turbulent field was recognized. In \cite{balkovsky01},
the case of particle with Stokes time in the turbulent viscous range was analyzed exploiting
the fact that, in this case, the fluid velocity is spatially smooth on the scale of interest 
for the solid particle. In both \cite{elperin00} and \cite{balkovsky01}, however, the inertial
range structure of the turbulent flow was disregarded altogether. The approach carried on here, 
allows instead to analyze the production of concentration fluctuations in any regimes of Stokes 
times, in particular in the inertial range, where qualitatively different behaviors for the 
concentration fluctuation build-up are observed. 

\vskip 5pt
Purpose of this paper is to extend the Kraichnan model to short but finite correlation 
times, preserving, in a controlled perturbation theory, the Lagrangian structure of correlations,
and providing several applications to the transport of particles with and without inertia.
The analysis will be confined to a situation of two-dimensional, stationary, homogeneous 
and isotropic turbulence.

This paper is organized as follows. In section II, the equations determining the extension of
the Kraichnan model will be illustrated and their main properties discussed. Section III
will be devoted to the dynamics of passive tracers;
the self-diffusion and relative diffusion of fluid parcels, including the expression
for the constants involved, will be determined;
the effect of finite diffusivity will be discussed
and the Batchelor constant for a passive scalar injected at large scale in the flow will 
be calculated.
In section IV, the transport properties of a heavy particle with Stokes time longer than the 
Kolmogorov time will be studied, focusing on the relation between the
correlation time for the fluid velocity sampled by the particle, and its Lagrangian and
Eulerian counterparts. Section V will be devoted to calculation of the concentration 
fluctuations arising from compressibility of the heavy particle flow. In section VI, 
the bias introduced by inertia in the sampling of fluid velocity by solid particles
(non-ergodic effects) will be analyzed. Section VII will be devoted to conclusions.

\vskip 10pt

\centerline{\bf II. Finite correlation time extension of the Kraichnan model}
\vskip 5pt
A two-dimensional random velocity field with Eulerian correlation times scaling like the 
eddy turn-over time of a real turbulent flow can be obtained very simply, writing 
appropriate Langevin equations for the Fourier components of the vorticity field:
$$
\partial_tq_\k(t)+\gamma_kq_\k(t)=h_k\xi_\k(t)
\eqno(2.1)
$$
where $q_\k(t)$ is the (space) Fourier transform of the vorticity:
$$
q(\x,t)=\nabla_\perp\cdot\u(\x,t),
\eqno(2.2)
$$
$[$for the generic vector ${\bf v}$, we indicate ${\bf v}_\perp=(-v_2,v_1)]$,
and $\xi_\k(t)$ is the Fourier transform of a zero mean fully uncorrelated noise term of 
unitary amplitude:
$$
\langle\xi(\x,t)\xi(0,0)\rangle=\delta(\x)\delta(t)
\eqno(2.3)
$$
The damping and forcing Kernels $\gamma_k$ and $h_k$ are chosen, for $k\ll\eta^{-1}$, 
with $\eta$ the Kolmogorov length of the flow, as:
$$
\gamma_k=\rho\Ckol^\frac{1}{2}\flux^\frac{1}{3}(k^2+k_0^2)^\frac{1}{3}
\eqno(2.4)
$$
$$
H_k=|h_k|^2=\frac{8\pi\rho\Ckol^\frac{3}{2}\flux k^2}{k^2+k_0^2}
\eqno(2.5)
$$
while, for $k\eta>1$, some cut-off is imposed on the forcing amplitude $H_k$.
In this way, the velocity
spectrum $\U_\k(t)$, defined by
$\langle\u_\k(t)\u_\p(0)\rangle=\U_\k(t)(2\pi)^2$ $\delta(\k+\p)$, will
read, for $k\eta\ll 1$:
$$
\U_\k(t) =4\pi\Ckol\flux^\frac{2}{3}\frac{\k_\perp\k_\perp}{k^2}
\frac{\exp(-\gamma_k|t|)}{(k^2+k_0^2)^\frac{4}{3}}
\eqno(2.6)
$$
where $\Ckol$ and $\flux$ play the role, respectively, of the Kolmogorov constant and the
inertial range energy flux in a real turbulent field having this correlation spectrum.
For $k_0\ll k\ll\eta^{-1}$, we thus have the energy spectrum:
$E_k=\Ckol\flux^\frac{2}{3}k^{-\frac{5}{3}}$.
Identifying $\gamma_k^{-1}$ with the decay time and $k^{-2}U_k^{-\frac{1}{2}}(0)$ with the
turn-over time of an eddy at scale $k^{-1}$, we see that $\rho$ gives the ratio of the 
eddy turn-over and eddy decay time in the inertial range. The effect of sweep by the large
scales, however, is not accounted for in this way.

The most natural way to impose Lagrangian correlations in the random velocity field is to include
an advection term in Eqn. (2.1), which will take the form in real space: 
$$
(\partial_t+\u(\x,t)\cdot\nabla)q(\x,t)+\int\d^2y\gamma(\x-\y)q(\y,t)=\int\d^2yh(\x-\y)\xi(\y,t)
\eqno(2.7)
$$
This has the form of a vorticity equation in which the forcing and dissipation term, instead
of being localized respectively at large and small scales, act over the whole of the inertial
range, and this is reflected in their being nonlocal operators in real space.
This is opposite to what happens in a real turbulent field, where energy balance is 
established between large scale forcing and small scale viscous dissipation, by means of 
the nonlinear cascade. A nonlinear cascade is still present because of the convection term,
but it acts on the timescale of the eddy turn-over time, and, for large $\rho$, its effect 
is only a correction to that of the forcing and damping terms.  Choosing $\rho$ large has
therefore the consequence that convection acts merely as a large scale sweep.

Actually, Eqn. (2.7) looks a lot like the typical starting point of many turbulent closures
\cite{kraichnan71,orszag77,yakhot86}, in 
which $\gamma_k$ gives the turbulent response function (eddy viscosity of small scales)
and $h_k$ the nonlinear forcing by the cascade. For instance, $\rho^{-2}$ coincides with the 
renormalized dimensionless coupling 
constant of the Renormalization Group (RNG) closure \cite{yakhot86,olla91}, and its smallness 
is there the basis for the establishment of a perturbation theory. Here, the philosophy is
rather different: no parametrization of the turbulence cascade is sought, $\rho$ is chosen
arbitrarily large, and the similarity with real turbulence is expected to be only kinematic.
(Also, the separation of a Kolmogorov constant out of the energy flux $\bar\epsilon$
is arbitrary).

Things can be made a little bit more quantitative, introducing scale by scale the sweep time:
$$
T_k=k^{-1}\langle u^2\rangle^{-\frac{1}{2}}
\sim \Ckol^{-\frac{1}{2}}\flux^{-\frac{2}{3}}k_0^\frac{1}{3}k^{-1}.
\eqno(2.8)
$$
i.e. the time needed to a vortex of size $k^{-1}$ to pass in front of a fixed probe.
We see that 
sweep is important for all scales for which $\gamma_kT_k<1$, i.e., from Eqns. (2.4) and 
(2.6), for $k>k_0\rho^3$.  The Kraichnan model is 
recovered when sweep can be neglected in all of the inertial range,
i.e. for $\rho>(\eta k_0)^{-\frac{1}{3}}$. This means basically that the zero correlation 
time limit is taken before the infinite Reynolds number limit $\eta k_0\to 0$.
In this regime we have:
$$
\U_\k(t)
\simeq \frac{\k_\perp\k_\perp}{k^2}\frac{2\pi}{\rho}\Ckol^\frac{1}{2}\flux^\frac{1}{3} 
k^{-\frac{10}{3}} \delta(t)
\eqno(2.9)
$$
To understand what happens in the regime of dominant sweep, it is convenient to shift
to Lagrangian coordinates. Introduce then the coordinate $\z(t|\x,t_0)$ of a fluid parcel
which at time $t_0$ is at $\x$, and define the Lagrangian velocity:
$$
\u^\smalL(\x,t)=\u(\z(t|\x,0),t)
\eqno(2.10)
$$
and analogous expressions for $q^\smalL(\x,t)$ and the other fields.
After introducing the increase of trajectory separation in a time $t$:
$\delta\z(t|\x,\y)=\z(t|\x,0)-\z(t|\y,0)-(\x-\y)$
Eqn. (2.7) becomes, in the new variables: 
$$
\partial_tq^\smalL(\x,t)+\int\d^2y\gamma(\x-\y+\delta\z(t|\x,\y))q^\smalL(\y,t)=
\int\d^2y h(\x-\y+\delta\z(t|\x,\y))\xi(\y,t)
\eqno(2.11)
$$
which must be coupled with the equation for $\delta\z$; inverting Eqn. (2.2):
$$
\partial_t\delta\z(t|\x,\y)=\frac{1}{2\pi}\int\d^2r[{\bf G}(\x,\r)-{\bf G}(\y,\r)]q^\smalL(\r,t)
\eqno(2.12)
$$
with 
$$
{\bf G}(\x,\r)=
\frac{(\x-\r+\delta\z(t|\x,\r))_\perp}
{|\x-\r+\delta\z(t|\x,\r)|^2}
\eqno(2.13)
$$
We see then that the natural expansion parameter of the theory is:
$$
\frac{\delta z(t|\x,\y)}{|\x-\y|}
\sim\frac{|\u(\x,0)-\u(\y,0)|}{|\x-\y|\gamma_{|\x-\y|^{-1}}}
\sim\rho^{-1}
\eqno(2.14)
$$
i.e. the relative amount of particle separation increase in an eddy lifetime. 
The zero order of the theory, which is Gaussian and is described by Eqn. (2.6) after 
substituting $\u\to\u^\smalL$, corresponds to neglecting trajectory 
separation in an eddy lifetime, while keeping the uniform large scale sweep, implicit 
in the Lagrangian field $q^\smalL$.

Although the results which follow in the present paper are all obtained to the lowest 
order in the $\rho$ expansion \cite{note1}, associated with neglecting all non-Gaussian effects in 
$\u$, a diagrammatic expansion of Eqns. (2.12-13) in terms of the fields $q^\smalL$, 
$\delta\z$ and their conjugate could be obtained by means 
of the Martin-Siggia-Rose formalism \cite{martin73}. 
This expansion would only be valid 
locally around $t=0$, since, at long times, trajectory separation
becomes dominant. $[$To be consistent, this perturbation expansion should not receive 
contribution by correlations involving pairs of points in space-time such that 
$\gamma_{|\x-\x'|^{-1}}|t-t'|>\rho$, but this is expected to be true from the exponential decay
of the time correlations$]$.

The interaction terms in the perturbation expansion are obtained Taylor expanding
the kernels $\gamma$, ${\bf G}$ and $h$ ($H$ working with the field action). 
The result for $\gamma$ is, for instance:
$$
\gamma(\x-\y+\delta\z(t|\x,\y))=\gamma(\x-\y)+\sum_{n=1}^\infty\lambda_{\gamma_n}
\gamma_n^{i_1...i_n}(\x-\y)\delta z_{i_1}(t|\x,\y)...\delta z_{i_n}(t|\x,\y)
\eqno(2.15)
$$
with $\lambda_{\gamma_n}=1$ a coefficient which may scale when carrying on power counting.
Similar coefficients $\lambda_{G_n}$ and $\lambda_{H_n}$ are introduced in the Taylor expansion
for $G$ and  $H$. The theory is thus characterized by an infinite number of interactions 
involving vertices, which, to $\O(\rho^{-n})$, have up to $2+n$ legs.

To check for divergences at large $k$ in the perturbation expansion, we use power
counting directly in Eqns. (2.11-13) \cite{zinn-justin}. Rescaling coordinates and times as:
$$
x\to\Lambda x\quad{\rm and}\quad t\to\Lambda^\frac{2}{3}t,
\eqno(2.16)
$$
Eqns. (2.9-11) remain invariant in form provided we rescale the various fields and interactions
$A=q^\smalL,\delta z$,
$\lambda_{\gamma_n},\lambda_{G_n},\lambda_{H_n}$:
as $A\to \Lambda^{[A]}A$, with
$$
[q^\smalL]=-\frac{2}{3}\qquad [\delta\z]=1\qquad
$$
$$
[\lambda_{\gamma_n}]=[\lambda_{H_n}] =[\lambda_{G_n}] =0.
\eqno(2.17)
$$
This leads to expect logarithmic divergences at large $k$, meaning renormalizability
of the field theory and the possibility of 
logarithmic correction to scaling, produced by renormalization of the parameters in 
Eqns. (2.11-13).

It must be mentioned that marginal interactions and renormalizability are consequence
of the dimensional relation implicit in Kolmogorov scaling:
$[q^\smalL]=-[t]$. In general, had we set:
$$
\gamma_k\sim k^r\qquad H_k\sim k^s
\eqno(2.18)
$$
we would have obtained:
$$
[q^\smalL]=\frac{r-s-2}{2}\qquad [\delta\z]=\frac{3r-s}{2}
$$
$$
[\lambda_{\gamma_n}]=[\lambda_{H_n}]
=[\lambda_{G_n}] =-n([t]+[q^\smalL]) 
=\frac{n}{2}(s+2-3r)
\eqno(2.19)
$$
We thus see that super-renormalizability $[\lambda]<0$ and non-renormalizability 
$[\lambda]>0$ of the theory occur, respectively, for positive and negative
$[q^\smalL]+[t]$ \cite{zinn-justin}.  This corresponds to the two regimes
of eddy decay 
time becoming asymptotically longer (shorter) than the eddy turn-over time, and hence 
the nonlinearity becoming dominant (negligible) at large scales.

Marginality of the interactions means that logarithmic divergences may arise both at
large and small $k$. At small $k$, however, such divergences are not expected, due 
to the subtraction in the definition of $\delta\z$. The reason is sketched below
(more details in a separate publication; it must be said, anyway, that this is not a surprise:
Lagrangian closures \cite{kraichnan65} were introduced precisely to cure the infrared divergences
arising in the original Eulerian theories). As it appears from Eqn. (2.17), small $k$ 
divergence is due to internal lines in a loop diagrams involving the field $\delta\z$.
The scaling of Eqn. (2.17) is associated with large, not with small $k$ behaviors. 
In fact, the
divergences occurring for large $k$ in a loop diagram will not change if we exchange 
$\delta\z(t|\x,\y)\to\z(t|\x,0)+\z(t|\y,0)-(\x+\y)$; this because each small eddy 
contribute to the separation $\x-\y$ an amount which is of the same order of 
the one to sweep.  Now, the logarithmic divergence predicted at small $k$ in a loop
diagram comes indeed, from equating the scaling of the sweep $\z(t|\x,0)+\z(t|\y,0)-(\x+\y)$ 
with that of the trajectory separation $\delta\z(t|\x,\y)$ also at small $k$, which 
is incorrect. For small $k$, this scaling should be corrected 
by a factor $k$ per field $\delta\z$ involved in the lines of the loop, 
and this is enough to eliminate divergence.

\vskip 10pt

\centerline{\bf III. Passive tracer transport}
\vskip 5pt
\noindent{\bf 1. Self-diffusion of a fluid parcel}
\vskip 5pt
Lagrangian correlation functions in the form 
$\langle\u^\smalL(\x,t)\u^\smalL(\x,0)\rangle=
\langle\u(\z(t|\x,0),t)\u(\x,0)\rangle$
are the simplest objects one may try to calculate from the random velocity field
introduced in section II. The starting point, to lowest order in $\rho^{-1}$, 
and after sending the Kolmogorov scale $\eta$ to zero,
is the following modification of Eqn. (2.6):
$$
\U^\smalL_\k(t)=
4\pi\Ckol\flux^\frac{2}{3}\frac{\k_\perp\k_\perp}{k^2}
\frac{\exp(-\gamma_k|t|)}{(k^2+k_0^2)^\frac{4}{3}}
\eqno(3.1)
$$
The Lagrangian correlation time $\tau_L$ is then readily calculated:
$$
\tau_L^{-1}=
\langle|\u^\smalL|^2\rangle
\Big[\int\d t\langle\u^\smalL(\x,t)\cdot\u^\smalL(\x,0)\rangle\Big]^{-1}=
2\rho\Ckol^\frac{1}{2}\flux^\frac{1}{3}k_0^\frac{2}{3}
\eqno(3.2)
$$
and we have the following relation between the turbulence level
$u_T^2=\langle u^2\rangle=\langle |\u^\smalL|^2\rangle$
and the integral scales of the flow $k_0$ and $\tau_L$:
$$
u_T^2
=3\Ckol\flux^\frac{2}{3}k_0^{-\frac{2}{3}}=6\rho\Ckol^\frac{3}{2}\flux\tau_L
\eqno(3.3)
$$
The correlation time $\tau_L$ is determined by the particular form of $\U^\smalL_\k$ we have
chosen at small $k$, which is non-universal. It is more interesting, and relevant from 
the point of view of Lagrangian dispersion modeling \cite{thomson87,sawford91}, to calculate 
the Lagrangian time structure function:
$$
\langle[u_\alpha^\smalL(\x,t)-u_\alpha^\smalL(\x,0)]
[u_\beta^\smalL(\x,t)-u_\beta^\smalL(\x,0)]\rangle
=\frac{1}{2}\langle|\u^\smalL(\x,t)-\u^\smalL(\x,0)|^2\rangle\delta_{\alpha\beta}
\eqno(3.4)
$$
We discover immediately that, in order to have a self similar spectrum for the inertial
range, the time correlations should have continuous time derivative at $t=0$, a property
not satisfied by Eqn. (3.1).

This self-similarity violation can be illustrated in a simple way imagining the turbulence
field in the neighborhood of the fluid parcel as a superposition of nested eddies with 
scale $l_n$, velocity $u_n$ and eddy turn-over time $\tau_n$:
$$
l_n=l_02^{-n},\qquad u_n=u_02^{-\frac{n}{3}},\qquad\tau_n=\tau_02^{-\frac{2n}{3}}
\eqno(3.5)
$$
If the time correlation decayed linearly for $t\to 0$, we would have:
$$
\langle |\u^\smalL(\x,t)-\u^\smalL(\x,0)|^2\rangle
\sim\sum_{\tau_n<t}u_n^2\frac{t}{\tau_n}+\sum_{\tau_n>t}u_n^2
\sim u_0^2\log(\tau_0/t)\frac{t}{\tau_0};
\eqno(3.6)
$$
Thus, identical scaling of $u_n^2$ and $\tau_n$, and linear decay of correlations
cause the largest space scale to contribute to the structure function at arbitrary
short time separation $t$, in the same way as a vortex with eddy turn-over
time $\tau_n\sim t$, whence the logarithmic correction involving $\tau_0$.

In order to have a quadratic behavior of the time correlation at $t=0$, it is 
necessary that the noise $\xi$ in Eqn. (2.7) be correlated in time, and the
correlation must again be given along the trajectories. The appropriate
modification to Eqn. (2.7) is therefore:
$$
\begin{cases}
(\partial_t+\u(\x,t)\cdot\nabla)q(\x,t)+\int\d^2y\gamma(\x-\y)q(\y,t)=r(\x,t)
\\
(\partial_t+\u(\x,t)\cdot\nabla)r(\x,t)+\int\d^2y\hat\gamma(\x-\y)r(\y,t)=\int\d^2yh(\x-\y)\xi(\y,t)
\end{cases}
\eqno(3.7)
$$
where, for $k\ll\eta^{-1}$:
$$
\hat\gamma_k=\hat\rho\Ckol^\frac{1}{2}\flux^\frac{1}{3}k^\frac{2}{3}
\quad{\rm and}\quad
H_k=|h_k|^2=8\pi\rho\hat\rho(\rho+\hat\rho)\Ckol^\frac{5}{2}\flux^\frac{5}{3}
(k^2+k_0^2)^\frac{2}{3}
\eqno(3.8)
$$
It is easy to show that also the field theory associated with Eqn. (3.7) is characterized
by marginal interactions: 
$[\lambda_{\gamma_n}]=[\lambda_{\hat\gamma_n}]=[\lambda_{H_n}]=[\lambda_{G_n}]=0$
and the considerations in section II extend to the present case.

The zero order of the theory leads to the following correlation function:
$$
\U^\smalL_\k(t)
=\frac{\k_\perp\k_\perp}{k^2}
\frac{4\pi\Ckol\flux^\frac{2}{3}}{(k^2+k_0^2)^\frac{4}{3}}
\frac{\rho\ex^{-\hat\gamma_k|t|}-\hat\rho\ex^{-\gamma_k|t|}}{\rho-\hat\rho}
\eqno(3.9)
$$
and the time correlation has a quadratic maximum at $t=0$.
Calculation of the Lagrangian correlation time leads to the same result of Eqn. (3.2), with 
the substitution $\rho\to\frac{\rho\hat\rho}{\rho+\hat\rho}$, while smoothness of the time
correlation eliminates the logarithmic correction to the scaling of the Lagrangian time
structure function. This structure function obeys in fact, after sending $k_0\to 0$,
the expected normal diffusion behavior: 
$$
\langle|\u^\smalL(\x,t)-\u^\smalL(\x,0)|^2\rangle=2C_0\flux|t|
\qquad
{\rm with} 
\qquad
C_0
=\Ckol^\frac{3}{2}\frac{\hat\rho\rho}{\hat\rho-\rho}\log\hat\rho/\rho;
\eqno(3.10)
$$
the constant $C_0$ is $\O(\rho)$ and,
as expected from the discussion leading to Eqn. (3.6),
diverges logarithmically for $\hat\rho/\rho\to\infty$.

\vskip 10pt
\noindent{\bf 2. Relative diffusion}
\vskip 5pt
Analyzing the transport of a cluster of particles requires consideration of time intervals,
during which the space separations involved cannot be approximated as constant. 
Over these timescales, the short correlation time limit leads to a perturbation scheme,
which treats the velocity field to zero order as a white noise.

We focus on the case of a pair of particles. We have to study an equation in the
form:
$$
\partial_t[z_\alpha(t|\r_0,0)-z_\alpha(t|0,0)]
=u_\alpha^\smalL(\r_0,t)-u_\alpha^\smalL(0,t)=U_{\alpha\beta}(\z(t|\r_0,0)-\z(t|0,0))
\xi_\beta(t)
\eqno(3.11)
$$
with $\langle\xi_\alpha(t)\xi_\beta(0)\rangle=\delta_{\alpha\beta}\delta(t)$ and
$U_{\alpha\beta}$ to be determined.
Due to the multiplicative noise nature of this equation, attention must be paid to the
possible presence of drift terms arising from the Stratonovich prescription implicit in
its definition \cite{gardiner}. It is easy to show that this drift is identically zero, either
by direct calculation of the increment $\delta\z(t|\x,0)$ for $t$ in the inertial
range, or noticing that:
$$
\langle\delta\z(t|\r_0,0)\rangle=\int_0^t\d\tau[
\langle (\u^\smalL(\r_0,\tau)\rangle-\langle\u^\smalL(0,\tau)\rangle]=0;
\eqno(3.12)
$$
this, because of homogeneity of turbulence. For this reason, the separation process is
described simply by:
$$
\partial_t\langle[z_\alpha(t|\r_0,0)-z_\alpha(t|0,0)][z_\beta(t|\r_0,0)-z_\beta(t|0,0)]\rangle=
D_{\alpha\beta}(\z(t|\r_0,0)-\z(t|0,0))
\eqno(3.13)
$$
with
$$
D_{\alpha\beta}(\r)=\int\d t\langle[u_\alpha^\smalL(\r,t)-u_\alpha^\smalL(0,t))]
[u_\beta^\smalL(\r,0)-u_\beta^\smalL(0,0))]\rangle
\eqno(3.14)
$$
This tensor is easily calculated from $D_{11}(\r)$ for $\r=(r,0)$,
exploiting incompressibility. Using $\int_0^{2\pi}\d\theta\sin^2\theta\sin^2(x\cos\theta)=
\frac{\pi}{2}[1-\J_0(2x)-\J_2(2x)]$, we find, in the limit $k_0\to 0$:
$$
D_{11}(\r)=\frac{4\alpha_\smalsev\Ckol^\frac{1}{2}\flux^\frac{1}{3}}{\rho}r^\frac{4}{3}
\eqno(3.15)
$$
where
$$
\alpha_\smalsev=\int_0^\infty\d x\  x^{-\frac{7}{3}}[1-\J_0(x)-\J_2(x)]
\simeq 0.265,
\eqno(3.16)
$$
with $\J_n$ the Bessel function of the first kind, is evaluated in terms of Gamma functions
\cite{gradshteyn} using the formula $\int_0^\infty\d x\ x^\mu J_\nu(x)=2^\mu
\frac{\Gamma(\frac{1}{2}(1+\nu+\mu))}{\Gamma(\frac{1}{2}(1+\nu-\mu))}$. From incompressibility we find therefore:
$$
D_{\alpha\beta}(\r)=\Big[\frac{r_\alpha r_\beta}{r^2}+\frac{7}{3}\Big(\delta_{\alpha\beta}
-\frac{r_\alpha r_\beta}{r^2}\Big)\Big]
\frac{4\alpha_\smalsev\Ckol^\frac{1}{2}\flux^\frac{1}{3}}{\rho}r^\frac{4}{3}
\eqno(3.17)
$$
We want to study the asymptotics of the separation process of two particles in the
inertial range. The procedure is standard (see e.g \cite{borgas94}); we introduce
the distribution $P$ for the
separation $\r$ at time $t$, which will obey the diffusion equation (the summation over
repeated indices convention is adopted throughout the paper):
$$
\partial_tP=\frac{1}{2}\partial_\alpha\partial_\beta D_{\alpha\beta}P
\eqno(3.18)
$$
and look for an isotropic similarity solution in the form
$$
P(\r,t)=t^{-3}f(t^{-\frac{3}{2}}r)=t^{-3}f(R)
\eqno(3.19)
$$
Equation (3.18) takes then the form
$$
\frac{3}{2}\partial_\alpha(R_\alpha f)
+\frac{4\alpha_\smalsev\Ckol^\frac{1}{2}\flux^\frac{1}{3}}{2\rho}\partial_\alpha R^\frac{1}{3}
R_\alpha\partial_Rf=0
\eqno(3.20)
$$
This equation has an unphysical solution, which is divergent in $R=0$, and a finite one:
$$
f(R)=\exp\Big(-\frac{9\rho R^\frac{2}{3}}{8\alpha_\smalsev\Ckol^\frac{1}{2}\flux^\frac{1}{3}}\Big)
\eqno(3.21)
$$
whose moments are:
$$
\langle R^n\rangle=\int_0^\infty R^{1+n}\d Rf(R)=\frac{3}{2}
\Big(\frac{8\alpha_\smalsev\Ckol^\frac{1}{2}\flux^\frac{1}{3}}{9\rho}\Big)^{3+\frac{3n}{2}}
\Gamma(3+\frac{3n}{2})
\eqno(3.22)
$$
with $\Gamma$ the standard gamma function.

From here, the expression for the particle space separation is obtained in a straightforward
manner; for $\gamma_{x^{-1}}t\gg 1$, indicating $\r(t)=\z(t|\r_0,0)-\z(t|0,0)$:
$$
\langle r^2(t)\rangle=c\flux t^3,
\qquad
c=\frac{10240\alpha_\smalsev^3\Ckol^\frac{3}{2}}{243\rho^3}
\eqno(3.23)
$$
i.e. the space separation obeys Richardson diffusion.
For the relative velocity, we have,
from Eqn. (2.6):
$$
\langle [u^\smalL_r(\r_0,0)-u^\smalL_r(0,0)]^2\rangle=2\alpha_\smalfi\Ckol\flux^\frac{2}{3}
\langle r^\frac{2}{3}(t)\rangle
\eqno(3.24)
$$
where
$$
\alpha_\smalfi=\int_0^\infty\d x\ x^{-\frac{5}{3}}[1-\J_0(x)-\J_2(x)]
\simeq 2.149
\eqno(3.25)
$$
and, using Eqn. (3.22), for $\gamma_{x^{-1}}t\gg 1$, we find the normal diffusion behavior:
$$
\langle [u^\smalL_r(\r_0,0)-u^\smalL_r(0,0)]^2\rangle=\tilde c\flux t,
\qquad
\tilde c=\frac{16\alpha_\smalfi\alpha_\smalsev\Ckol^\frac{3}{2}}{3\rho}
\eqno(3.26)
$$
Passing to the smoothed out in time version of the velocity field provided by Eqn. (3.7),
is accomplished, as in the case of $\tau_L$, by exchanging
$\rho\to\frac{\rho\hat\rho}{\rho+\hat\rho}$.
In \cite{boffetta00}, both a sub-exponential
behavior for the function $f(R)$ and Richardson diffusion were observed in a direct DNS
(direct numerical simulation) of two-dimensional turbulence in the inverse cascade regime. 
Based on the results of that paper, extrapolating applicability of our leading order 
expressions in $\rho$ would give then (taking also $\hat\rho\to\infty$): $\rho\simeq 2$.

\vskip 10pt
\noindent{\bf 3. The role of diffusivity and the Batchelor constant}
\vskip 5pt
The dynamics of passive tracers, contrary to that of fluid elements, feels the effect of 
molecular diffusivity. Due to finiteness of the turbulent 
correlation times, this effect does not consist purely of an additive noise contribution 
to the tracer velocity. Indicating by $\sigma$ the molecular diffusivity,
the passive tracer velocity will have the form:
$$
\v(\x,t)+(2\sigma)^\frac{1}{2}{\boldsymbol\xi}(\x,t)
\eqno(3.27)
$$ 
with $\langle\xi_\alpha(\x,t)\xi_\beta(0,0)\rangle=\delta_{\alpha\beta}\delta(\x)\delta(t)$ and
$\v$ obeying an equation in the form:
$$
(\partial_t+\v(\x,t)\cdot\nabla)q_v(\x,t)+\int\d^2y\gamma(\x-\y)q_v(\y,t)
-\int\d^2yh(\x-\y)\xi(\y,t)
$$
$$
=-(2\sigma)^\frac{1}{2}\langle{\boldsymbol\xi}(\x,t)\cdot\nabla q_v(\x,t)\rangle_\xi
\simeq\sigma\nabla^2q_v(\x,t)
\eqno(3.28)
$$
where $q_v=\nabla_\perp\cdot\v$, $\langle .\rangle_\xi$ is an average limited to the 
noise ${\boldsymbol\xi}$ and 
use has been made, in converting the advection by molecular noise into a diffusion term, 
of It\^o's lemma \cite{gardiner}.  We see (it is assumed that the limit $\eta\to 0$ is already 
taken) that there is a renormalization of the damping kernel $\gamma$:
$$
\gamma_k\to\gamma_k+\sigma k^2
\eqno(3.29)
$$
which leads to a cut-off for the velocity at the inverse diffusive scale
$$
\eta^{-1}_\sigma=(\rho\Ckol^\frac{1}{2})^\frac{3}{4}\flux^\frac{1}{4}\sigma^{-\frac{3}{4}}
\eqno(3.30)
$$
We have then:
$$
\langle \v_\k(t)\v_{-\k}(0)\rangle
=\frac{\exp(-\sigma k^2|t|)}{1+(k\eta_\sigma)^\frac{4}{3}}\langle \u_\k(t)\u_{-\k}(0)\rangle,
\eqno(3.31)
$$
and for small space separations $r/\eta_\sigma\to 0$, we have a quadratic behavior for the 
velocity structure function:
$$
\langle[v_r(\x+\r,t)-v_r(\x,t)]^2\rangle
=2\Ckol\flux^\frac{2}{3}r^2\eta_\sigma^{-\frac{4}{3}}
\int_0^\infty\frac{[1-\J_0(x)-\J_2(x)]\d x}{x^\frac{5}{3}(x^\frac{4}{3}+(r/\eta_\sigma)^\frac{4}{3})}
$$
$$
\simeq \frac{1}{4}\Ckol\flux^\frac{2}{3}\eta_\sigma^{-\frac{4}{3}}r^2|\log r/\eta_\sigma |
\eqno(3.32)
$$
The transport of a passive scalar $\theta(\x,t)$ will be described by the equation
$$
(\partial_t+\v(\x,t)\cdot\nabla)\theta(\x,t)=\sigma\nabla^2\theta(\x,t)+f(\x,t)
\eqno(3.33)
$$
with $f(\x,t)$ a source term. An interesting quantity to calculate is the fluctuation 
spectrum for $\theta$ in the case $f$ is random in time and concentrated at large scale:
$$
\langle f(\x+\r,t)f(\x,0)\rangle=F(r)\delta(t),\qquad 
F(r)=
\begin{cases}
2\flux_\theta, & k_0r<1
\\
0, & k_0r>0
\end{cases}
\eqno(3.34)
$$
We can thus consider $\langle\theta\rangle=0$.
The equation for the steady state passive scalar correlation
$\Theta(r)=\langle\theta(\x+\r,t)\theta(\x,t)\rangle$ will then be, for $k_0r\ll 1$:
$$
\langle\theta(\x,t)[\v(\x+\r,t)-\v(\x,t)]\cdot\nabla\theta(\x+r,t)\rangle
=2\sigma\nabla^2\Theta(r)+4\flux_\theta
\eqno(3.35)
$$
For $r\to 0$, the left hand side of this equation is zero; we thus obtain 
$\flux_\theta=\frac{\sigma}{2}\langle|\nabla\theta|^2\rangle$, i.e. $\flux_\theta$ is
the dissipation of passive scalar fluctuations.
Following the same approach of the previous section, the velocity difference 
$\v(\x+\r,t)-\v(\x,t)$ is approximated by a white noise. From It\^o's Lemma, its 
contribution in 
Eqn. (3.35) will be an eddy diffusivity $D_{\alpha\beta}^v(\r)$, whose expression
will coincide, for $r\gg \eta_\sigma r$,
with the one for $D_{\alpha\beta}$ provided by Eqns. (3.16-17).
Drift terms coming from the Stratonovich prescriptions are
ruled out with the same arguments used in the previous section.  
The resulting diffusion equation will then read: 
$$
\partial_\alpha\partial_\beta(\frac{1}{2}D_{\alpha\beta}^v(\r)+2\sigma\delta_{\alpha\beta})
\Theta(r)+4\flux_\theta=0
\eqno(3.36)
$$
For $r\gg\eta_\sigma$ $D^v_{ij}$ is essentially a correction to the
molecular diffusivity, and will read, from Eqn. (3.31):
$$
D_{\alpha}^v(\r)=
\int\d t\langle[v_\alpha(\r,t)-v_\alpha(0,t))][v_\beta(\r,0)-v_\beta(0,0))]\rangle
$$
$$
\simeq
\Big[\frac{r_\alpha r_\beta}{r^2}+\frac{13}{3}\Big(\delta_{\alpha\beta}-\frac{r_\alpha 
r_\beta}{r^2}\Big)\Big]
\frac{3\pi\sigma}{8\rho^2}(r/\eta_\sigma)^\frac{10}{3}
\eqno(3.37)
$$
For $\eta_\sigma\ll r$, $D^v_{ij}$ is approximated by Eqn. (3.17), the molecular
diffusivity $\sigma$ can be neglected and Eqn. (3.35) takes the form:
$$
\Theta''+\frac{7}{3r}\Theta'=-\frac{8\rho\flux_\theta}
{4\alpha_\smalsev\Ckol^\frac{1}{2}\flux^\frac{1}{3}r^\frac{4}{3}}
\eqno(3.38)
$$
Solution of this equation gives automatically the passive scalar
structure function $\langle[\theta(\x+\r,t)-\theta(\x,t)]^2\rangle=2[\Theta(0)-\Theta(r)]$
in the inertial range for $\theta$: $\eta_\sigma\ll r\ll k_0^{-1}$. This structure
function scales like $r^\frac{2}{3}$ and can be written in the form:
$$
\langle[\theta(\x+\r,t)-\theta(\x,t)]^2\rangle
=\frac{B\flux_\theta r^\frac{2}{3}}{\Ckol^\frac{1}{2}\flux^\frac{1}{3}}
\eqno(3.39)
$$
with the parameter $B=\frac{3\rho}{\alpha_\smalsev}$ the so called Batchelor constant 
of the flow. As with relative diffusion, the case of a velocity field with 
smooth time correlation described by Eqn. (3.7) is recovered substituting 
$\rho$ with $\frac{\rho\hat\rho}{\rho+\hat\rho}$.


\vskip 10pt
\centerline{\bf IV. Solid tracers: 1-particle statistics}
\vskip 5pt
We consider the simplest case of a linear drag. In the presence of gravity (or of a constant
external force) and of the
turbulent velocity field $\u(\x,t)$, the solid particle coordinate $\z^\smalP(t|\x,0)$
will obey the equation of motion:
$$
\dot\z^\smalP(t|\x,t)=\v^\smalP(\x,t)+\u_G,\qquad \z^\smalP(0|\x,0)=\x
\eqno(4.1)
$$
where $\u_G$ is the gravitational drift, that we suppose constant and uniform
and $\v^\smalP$ is the fluctuation in the Lagrangian solid particle velocity,
which obeys the linear relaxation equation:
$$
\dot\v^\smalP(\x,t)=\tau_S^{-1}(\u(\z^\smalP(t|\x,0),t)-\v^\smalP(\x,t))
=\tau_S^{-1}(\u^\smalP(\x,t)-\v^\smalP(\x,t)),
\eqno(4.2)
$$
with $\tau_S$ the Stokes time. (For a spherical particle of radius $a$ and density $\rho_\smalP$, in
a fluid of density $\rho_{\scriptscriptstyle{0}}$ and kinematic viscosity $\nu$, we would have:
$\frac{2a^2}{9\nu}|1-\rho_\smalP/\rho_{\scriptscriptstyle{0}}|$; we are disregarding
any effect from finite particle Reynolds number \cite{maxey83}).
From now on we shall identify Lagrangian quantities calculated on 
solid particle trajectories by the superscript $P$.

In general the non-coincidence of fluid and solid particle trajectories makes the analysis
of Eqns. (4.1-2) a very difficult task. The short correlation time limit $\rho\to\infty$,
however, allows to proceed perturbatively in the fluctuating part of the trajectory separation 
$\u_Gt+\z(t|\x,0)-\z^\smalP(t|\x,0)$. The physical motivation for this is that, from Eqn. (4.2),
$\u_Gt+\z(t|\x,0)-\z^\smalP(t|\x,0)$ fluctuates on timescale $\tau_S$ with velocity scale fixed 
by those eddies which have decay time $\tau_S$. Hence, for $\rho$ large, the fluctuating part 
of trajectory separation remains small on the scale of these eddies. Furthermore, when either
$u_Gt>\delta z(t|\x+\u_Gt,\x)$, or $\gamma_{|u_Gt|^{-1}}t> 1$, in other words, when either
$\frac{\Ckol^\frac{3}{2}\flux t}{u_G^2} < 1 $ or $\frac{\Ckol^\frac{3}{2}\flux t}{u_G^2}>\rho^{-3}$
(provided $\rho> 1$, one of the two conditions is always satisfied), 
it is possible to approximate $\z(t|\x,0)+\u_Gt\simeq \z(t|\x+\u_Gt,0)$.

To lowest order we have therefore: 
$$
\u(\z^\smalP(t|\x,0),t)=\u(\u_Gt+\z(t|\x,0),t)=\u^\smalL(\x+\u_Gt,t)
\eqno(4.3)
$$
We obtain immediately the fluctuation amplitude of the velocity difference between solid 
and fluid particle at a given position. From Eqns. (4.2-3) we can write:
$$
\v^\smalP(\x,t)=\int\frac{\d^2k}{(2\pi)^2}\int_{-\infty}^t\frac{\d\tau}{\tau_S}\u^\smalL_\k(\tau)
\exp(-\frac{t-\tau}{\tau_S}+\i\k\cdot\x)
\eqno(4.4)
$$
and from here we obtain, using Eqn. (3.2):
$$
\langle(v_\alpha-u_\alpha)(v_\beta-u_\beta)\rangle=\delta_{\alpha\beta}u^2_S\int_1^\infty\frac{\d x}
{x(1+\frac{2\tau_S}{\tau_L}x^\frac{1}{3})}
\underset{\scriptscriptstyle \tau_S\ll\tau_L}\longrightarrow 
3\delta_{\alpha\beta}u^2_S\log(\tau_L/\tau_S)
\eqno(4.5)
$$
where 
$$
u_S=\Big(\frac{\tau_S}{3\tau_L}\Big)^\frac{1}{2}u_T
\eqno(4.6)
$$
for $\tau_S<\tau_L$, is the velocity scale of eddies with lifetime $\tau_S$ and $u_T$ is the
turbulent velocity defined in Eqn. (3.3). In order to proceed to next order, it is necessary to 
calculate the trajectory separation:
$$
\z^\smalP(t|\x,0)-\z(t|\x,0)=\u_Gt+(1-\ex^{-t/\tau_S})
\int_{-\infty}^0\d\tau\ \ex^{\tau/\tau_S}\u^\smalL(\x,\tau)
$$
$$
-\int_0^t\d\tau\exp(-\frac{t-\tau}{\tau_S})\u^\smalL(\x,\tau)
\eqno(4.7)
$$
We notice from this equation that the inertia produced part of trajectory separation 
does not grow indefinitely. In other words, if $\u_G=0$ and to lowest order in $\rho^{-1}$, 
there will be localization of 
solid particle trajectories around the fluid parcel trajectories they cross at any given
time; from Eqn. (4.7): $\langle |\z^\smalP(t|\x,-\infty)-\z(t|\x,-\infty)|^2\rangle
\sim (u_T\tau_S)^2\sim \Ckol\flux^\frac{2}{3}k_0^{-\frac{2}{3}}\tau_S^2$.
We thus introduce the localization length $S_l$:
$$
S_l=\Ckol^\frac{1}{2}\flux^\frac{1}{3}k_0^{-\frac{1}{3}}\tau_S
\eqno(4.8)
$$
What happens is that the velocity difference $\v^\smalP-\u^\smalP$ obeys a relaxation equation
with a forcing which is a time derivative; from Eqn. (4.2):
$\frac{\d}{\d t}(\v^\smalP-\u^\smalP)+\tau_S^{-1}(\v^\smalP-\u^\smalP)=-\dot\u^\smalP$.
The frequency spectrum of $\v^\smalP-\u^\smalP$ does not have therefore the small frequency 
singularity necessary for long time divergence. 
The localization length $S_l$ will appear to play a fundamental role in the production 
both of concentration fluctuations and of corrections to the velocity correlation time.
(Of course, to higher order in $\rho^{-1}$, the relative separation of fluid parcels sets in  
and localization is destroyed; $S_l$ becomes then, that part of trajectory separation which 
remains after the Richardson diffusion contribution is subtracted out). 

In the absence of gravity, beside the integral scale dependent localization length $S_l$, 
three more scales,
which, if $\tau_S\ll\tau_L$, are purely inertial,
can be obtained combining $\tau_S$, the crossing time of an eddy by a solid particle,
the eddy lifetime and the eddy turn-over time. We have the size $S$ of an eddy whose
lifetime equals
$\tau_S$: $\gamma_{S^{-1}}\tau_S\sim 1$; the size $S_c$ of an eddy that is crossed by a solid
particle in a time $\tau_S$: $u_S\sim S_c/\tau_S$; the size $S_i$ of an eddy whose lifetime
equals the crossing time by a solid particle: $S_i\gamma_{S_i^{-1}}\sim u_S$. Summarizing:
$$
S=\rho^\frac{3}{2}\Ckol^\frac{3}{4}\flux^\frac{1}{2}\tau_S^\frac{3}{2},
\qquad
S_c=\rho^\frac{1}{2}\Ckol^\frac{3}{4}\flux^\frac{1}{2}\tau_S^\frac{3}{2}
\quad
{\rm and}
\quad
S_i=\rho^{-\frac{3}{2}}\Ckol^\frac{3}{4}\flux^\frac{1}{2}\tau_S^\frac{3}{2}
\eqno(4.9)
$$
From Eqn. (4.9), we identify the following sequence of ranges:
\begin{itemize}
\item A large separation range $r>S$, in which the fluid velocity $\u^\smalP$ varies slowly
on the scale of the relaxation time $\tau_S$.
\item A first intermediate range $S<r<S_c$ in which the fluid velocity $\u^\smalP$ is
a fast variable, but still, $\tau_S$ is short compared with the crossing time of an eddy of 
size $r$;
hence, Eqn. (4.2) has the form of a Langevin equation with a noise $\tau_S^{-1}\u^\smalP$
of constant amplitude on the scale of this crossing time. The cross-over scale $S$ will
play an important role in the determination of the degree of non-ergodicity of the solid particle
flow (see section VI).
\item A second intermediate range $S_c<r<S_i$, in which the crossing time is shorter than both
the Stokes time and the eddy turn-over time, but is longer than the lifetime of an eddy of that
size; hence, the solid particle moves ballistically with respect to the fluid.
\item A small separation range $r<S_i$, in which trajectory separation in the lifetime of an 
eddy, is not a perturbation any more. 
\end{itemize}

From Eqns. (4.3), (4.4) and (4.7), we can establish a perturbative calculation scheme
for $\u^\smalP$ and $\v^\smalP$. Notice that, within perturbation theory, $\v^\smalP$ is
a one-valued function of $\x$ and $t$, and $\v(\x,t)$ defines automatically a velocity
field for the solid particles. The separation between $S_i$ and all the other 
scales of the problem, has the consequence that, in the present case, the Weinstock
approximation is exact \cite{weinstock76}. What happens is that trajectory separation 
is produced mainly by eddies of size $r\gtrsim S$, for which
trajectory separation is a perturbation. This has the consequence, in particular, that
the Weinstock approximation applies also at scales $r<S_i$ for which trajectory separation 
is not a perturbation at all. For dominant gravity, i.e. when $u_G>u_S$, trajectory separation
is produced mainly by the gravitational drift $u_G$ and the Weinstock
approximation is automatically satisfied.

We can calculate at this point the time correlation for the solid particle velocity and
adopt the approach followed in \cite{pismen78,nir79}; 
we can thus write, using Eqn. (4.7):
$$
\langle u_1^\smalP(0,0)u_1^\smalP(0,t)\rangle
=\int\frac{\d^2k}{(2\pi)^2}\frac{\d^2p}{(2\pi)^2}
\langle  u^\smalL_{1\k}(0)u^\smalL_{1\p}(t)
\exp[\i\p\cdot(\z^\smalP(t|0,0)-\z(t|0,0))]\rangle
$$
$$
=-\int\frac{\d^2k}{(2\pi)^2}\frac{\d^2p}{(2\pi)^2}
\exp(\i\p\cdot\u_Gt)\frac{\delta^2Z[{\bf J}]}{\delta J_{\k 1}(0)\delta J_{\p 1}(t)}
\Big|_{{\bf J}=\p\bar J_t}
\eqno(4.10)
$$
where 
$$
Z[{\bf J}]=\Big\langle\exp\Big(\i\int\frac{\d^2s}{(2\pi)^2}
\int\d t\u^\smalL_\s(t)\cdot{\bf J}_\s(t)\Big)\Big\rangle
$$
$$
=\N\exp\Big(-\frac{1}{2}\int\d\tau\d\tau'\int\frac{\d^2s}{(2\pi)^2}
{\bf J}_\s(\tau)\cdot\U_\s^\smalL(\tau-\tau')\cdot{\bf J}_{-\s}(\tau')\Big)
\eqno(4.11)
$$
is the generating functional for the field $\u^\smalL$ and
$$
\bar J_t(\tau)=
\begin{cases}
0, &\tau>t
\\
-\exp(-\frac{t-\tau}{\tau_S}), & 0<\tau<t
\\
(1-\exp(-t/\tau_S))\exp(\tau/\tau_S), &\tau<0
\end{cases}
\eqno(4.12)
$$
Substituting back into Eqn. (4.10), we obtain, after introducing dimensionless
variables $\bar t=t/\tau_S$, $\bar u_G=k_0\tau_Su_G$ and $\bar\gamma=\tau_S\gamma_{k_0}=
\tau_S/(2\tau_L)$:

$$
\langle u_1^\smalP(0,0)u_1^\smalP(0,t)\rangle
=\frac{u_T^2}{6}
\int_1^\infty\d x\  x^{-\frac{4}{3}}[\J_0(\bar u_G(x-1)^\frac{1}{2}\bar t)+
\J_2(\bar u_G(x-1)^\frac{1}{2}\bar t)]
$$
$$
\times\exp\Big[-\bar\gamma \bar t x^\frac{1}{3}
-\frac{\bar\gamma^3(x-1)}{2\rho^2}
\int_1^\infty\frac{\d y}{y^\frac{4}{3}(1+\bar\gamma y^\frac{1}{3})}
\Big(1-\ex^{-\bar t}-\frac{\ex^{-\bar\gamma\bar ty^\frac{1}{3}}-\ex^{-\bar t}}
{1-\bar\gamma y^\frac{1}{3}}\Big)\Big]
\eqno(4.13)
$$
We see from this equation that decorrelation of the fluid velocity sampled 
by a solid particle receives three contributions: one from the gravitational 
drift $u_G$, one from the eddy decay $\bar\gamma x^\frac{1}{3}\bar t$ and the 
integral term in the exponential, which comes from inertia produced trajectory 
separation. This last term is peculiar, in that it saturates to a constant
for long $t$ instead of continuing to increase indefinitely. This term is the
argument in the exponential expression for $Z[\J]$ $[$see Eqn. (4.11)$]$, which 
is essentially:
$$
\p\p : \langle[\z^\smalP(t|0,0)-\z(t|0,0)][\z^\smalP(t|0,0)-\z(t|0,0)]\rangle
\eqno(4.14)
$$
with the drift $\u_G$ subtracted out, and with $\p$ the wavevector entering 
the integral of Eqn. (4.10). But, from Eqns. (4.7-8), we saw that this expression
saturates at $t\to\infty$. In consequence of this, for long 
enough times, the large $x$ behavior of the integrand in Eqn. (4.13) will be 
dominated by the value at saturation of the inertia produced term. 

\vskip 10pt
\noindent{\bf 1. Velocity self-diffusion}
\vskip 5pt
Inertia causes two ranges of time separations in the correlation 
$\langle u_1^\smalP(\x,0)u_1^\smalP(\x,t)\rangle$: one at short times dominated by sweep from 
the velocity difference $\u_G+\v-\u$ and one at long time associated with eddy decay,
where Eqn. (3.10) holds \cite{olla01}.  The transition between the two ranges occurs at
$$
t\sim \frac{\max(u_G^2,u_S^2)}{\rho^3\Ckol^\frac{3}{2}\flux}
\eqno(4.15)
$$
From Eqns. (4.5-6), for dominant inertia, i.e. $u_S>u_G$, this cross-over time is much 
shorter than $\tau_L$, while, for dominant gravity, i.e. for $u_G\gg u_S$ it is
possible that sweep dominates for all inertial timescales; for this to occur, it is
necessary that the crossing time of a large eddy by the particle be less than $\tau_L$,
i.e. $k_0u_G\tau_L>1$. For dominant inertia the cross-over time 
$u_S^2/(\rho^3\Ckol^\frac{3}{2}\flux)\sim\rho^{-2}\tau_S$ is just the lifetime of an 
eddy of size $S_i$ $[$see Eqn. (4.9)$]$.

For dominant gravity, the exponential term in Eqn. (4.13) can be neglected. For
$t\ll\min(\tau_G,\tau_L)$ with 
$\tau_G=\frac{6}{\rho^2}(\frac{u_G}{u_T})^2\tau_L\sim\frac{u_G^2}{\rho^3\Ckol^{3/2}\flux}$, 
we find:
$$
\langle[u_1^\smalP(0,t)-u_1^\smalP(0,0)]^2\rangle=
\frac{u_T^2}{3}
\int_1^\infty\d x\ x^{-\frac{4}{3}}[1-\J_0(\bar u_G\bar t(x-1)^\frac{1}{2})-
\J_2(\bar u_G\bar t(x-1))^\frac{1}{2})]
$$
$$
\simeq
\frac{2}{3}\alpha_\smalfi\Ckol\flux^\frac{2}{3}(u_Gt)^\frac{2}{3}
\eqno(4.16)
$$
where $\alpha_\smalfi\simeq 2.149$ $[$see Eqn. (3.25)$]$.
The time $\tau_G$, for $u_G<u_L$, is the lifetime of vortices whose lifetime equals
the crossing time by a falling particle; for $\rho=O(1)$, $\tau_G$ coincides with the eddy
turn-over time of vortices with characteristic velocity $u_G$. 

For dominant inertia $u_G<u_S$ and short enough times $t\ll\rho^{-2}\tau_S$, only the last piece in
Eqn. (4.13) will contribute and will be quadratic in $\bar t$;
if $\tau_S\ll\tau_L$:
$$
1-\ex^{-\bar t}-\frac{\ex^{-\bar\gamma\bar ty^\frac{1}{3}}-\ex^{-\bar t}}
{1-\bar\gamma y^\frac{1}{3}}
\simeq
\frac{1}{2}\bar\gamma y^\frac{1}{3}\bar t^2
\eqno(4.17)
$$
Substituting into Eqn. (4.13), we are left with the following expression:
$$
\langle[u_1^\smalP(0,t)-u_1^\smalP(0,0)]^2\rangle
=\frac{u_T^2}{3}
\int_1^\infty\d x\  x^{-\frac{4}{3}}
\Big\{1-\exp\Big[-\frac{\bar\gamma^3\bar t^2(x-1)}{4\rho^2}
\int_1^\infty\frac{\d y}{y(1+\bar\gamma y^\frac{1}{3})}
\Big]\Big\}
$$
$$
\simeq \frac{u_T^2}{3}
\int_0^\infty\d x\  x^{-\frac{4}{3}}
\Big\{1-\exp\Big[\frac{3\bar\gamma^3\bar t^2x\log\bar\gamma}{4\rho^2}
\Big]\Big\}
\eqno(4.18)
$$
where use has been made again, in passing, from the first to the second line,
of the condition $t\ll\rho^{-1}\tau_S$.
Using $\int_0^\infty\d x\ x^{-\frac{4}{3}}(1-\exp(-Ax))=3\Gamma(2/3)A^\frac{1}{3}$,
we obtain therefore:
$$
\langle[u_1^\smalP(0,t)-u_1^\smalP(0,0)]^2\rangle
\simeq
\frac{3}{2}\Gamma(2/3)\Big(3\log(\tau_L/\tau_S)\Big)^\frac{1}{3}
\Ckol\flux^\frac{2}{3}
(u_St)^\frac{2}{3}
\eqno(4.19)
$$
As predicted in \cite{olla01}, at short times, the time structure function for $\u^\smalP$ has a 
sub-diffusive behavior with exponent $\frac{2}{3}$ both for dominant $u_G$ and dominant $u_S$.
What happens is that at such short time scales, the particle crosses at constant
speed (remember also, in the inertia dominated case, that $S_c\gg S_i$) vortices whose velocity field is, 
in the limit, basically frozen; hence a Taylor hypothesis applies, and time  
correlations coincide with their spatial counterparts.


\vskip 10pt
\noindent{\bf 2. Velocity correlation times}
\vskip 5pt
Starting from Eqn. (4.13), we can calculate the correlation time $\tau_P$ for the fluid velocity 
sampled by a solid particle:
$$
\tau_P=\langle [u_1^\smalP]^2\rangle^{-1}\int_0^\infty\d t
\langle u_1^\smalP(0,0)u_1^\smalP(0,t)\rangle
\eqno(4.20)
$$
To lowest order, any discrepancy between the PDF (probability distribution functions) for
$u^\smalL$ and $u^\smalP$ can be neglected and we have $\langle [u_1^\smalP]^2\rangle
=\langle [u_1^\smalL]^2\rangle=\frac{1}{2}u_T^2$.
We begin by analyzing the case of dominant inertia: $u_G=0$. Taylor expanding in $\rho^{-1}$
the integrand in Eqn. (4.13) and substituting into Eqn. (4.20), leads to terms which diverge
when integrated in $x$. This indicates that the time independent part of the inertia term
in Eqn. (4.13) dominates the integral.  
We thus Taylor expand in $\rho^{-1}$,
only the time dependent piece of the integrand in Eqn. (4.13), i.e.:
$$
\exp\Big[-\bar\gamma\bar tx^\frac{1}{3}+\frac{\bar\gamma^3(x-1)}{2\rho^2}
\int_1^\infty\frac{\d y}{y^\frac{4}{3}(1+\bar\gamma y^\frac{1}{3})}
\Big(\ex^{-\bar t}+\frac{\ex^{-\bar\gamma\bar ty^\frac{1}{3}}-\ex^{-\bar t}}
{1-\bar\gamma y^\frac{1}{3}}\Big)\Big]
\eqno(4.21)
$$
to obtain:
$$
\int_0^\infty\d t\langle u_1^\smalP(\x,0)u_1^\smalP(\x,t)\rangle
=\frac{u_T^2}{6}
\int_1^\infty\d x\  x^{-\frac{4}{3}}
\exp\Big(-\frac{\bar\gamma^2(x-1)}{2\rho^2}
\int_1^\infty\frac{\d y}{y^\frac{4}{3}(1+\bar\gamma y^\frac{1}{3})}\Big)
$$
$$
\times\Big[\frac{1}{\bar\gamma x^\frac{1}{3}}+
\frac{\bar\gamma^2(x-1)}{2\rho^2}
\int_1^\infty\d y
\frac{1+\bar\gamma x^\frac{1}{3}-\bar\gamma^2 y^\frac{1}{3}(x^\frac{1}{3}+y^\frac{1}{3})}
{y^\frac{4}{3}(1-\bar\gamma^2y^\frac{2}{3})(1+\bar\gamma x^\frac{1}{3})
(x^\frac{1}{3}+y^\frac{1}{3})}\Big]
\eqno(4.22)
$$
and we see that the integral in $x$ of the $\O(\rho^{-2})$ on second line of Eqn. (4.22) is
dominated in fact by a saddle point at $x=(k/k_0)^2\sim (\rho/\bar\gamma)^2$, i.e. at 
$k\sim S_l^{-1}$. 
Combining this result, with the fact that the integrands are peaked at $y\sim 1$, 
Eqn. (4.22) will take the form:
$$
\int_0^\infty\d t\langle u_1^\smalP(0,0)u_1^\smalP(0,t)\rangle
=\frac{u_T^2}{6}
\int_1^\infty\d x\  x^{-\frac{4}{3}}
\exp\Big(-\frac{\bar\gamma^2x}{2\rho^2}
\int_1^\infty\frac{\d y}{y^\frac{4}{3}(1+\bar\gamma y^\frac{1}{3})}\Big)
$$
$$
\times\Big[\frac{1}{\bar\gamma x^\frac{1}{3}}+
\frac{\bar\gamma^2x^\frac{2}{3}}{2\rho^2}
\int_1^\infty\frac{\d y}{y^\frac{4}{3}(1+\bar\gamma y^\frac{1}{3})}\Big].
\eqno(4.23)
$$
We thus obtain, for the deviation $\tau_P-\tau_L$:
$$
\frac{\tau_P}{\tau_L}=1+B(\bar\gamma)\bar\gamma^\frac{4}{3}\rho^{-\frac{4}{3}}+\O(\rho^{-2})
\eqno(4.24)
$$
where
$$
B(\bar\gamma)=\Big(\frac{2}{3}\Big)^\frac{1}{3}\Gamma(1/3)
\Big[\frac{1}{3}-\frac{\bar\gamma}{2}+\bar\gamma^2+\bar\gamma^3\log\frac{\bar\gamma}{1+\bar\gamma}\Big]^\frac{2}{3}
\eqno(4.25)
$$
It is to be noticed that the factor $B(\bar\gamma)$ is always positive, i.e. the correlation 
time for the fluid velocity seen by the solid particle is longer than $\tau_L$. Following the
argument in \cite{kraichnan64a}, this would be expected in the case of a velocity field with 
statistics
defined in an Eulerian frame, and is exactly the result obtained in \cite{reeks77}.
In the case of a Lagrangian statistics, it is not clear whether 
the deviation between solid and fluid particle trajectories, should have lead to 
a faster, rather than slower decorrelation rate. 

In the case of dominant gravity, as expected \cite{csanady63,nir79}, there is always 
a decrease of the correlation time. In place of Eqn. (4.22), we have:
$$
\int_0^\infty\d t\langle u_1^\smalP(0,0)u_1^\smalP(0,t)\rangle
$$
$$
=\frac{u_T^2}{6}
\int_0^\infty\d t\int_1^\infty\d x\  x^{-\frac{4}{3}}
[\J_0(\bar u_G(x-1)^\frac{1}{2}\bar t)+\J_2(\bar u_G(x-1)^\frac{1}{2}\bar t)]
\exp(-\bar\gamma \bar t x^\frac{1}{3})
\eqno(4.26)
$$
which, using $\int_0^\infty\d x\J_\nu(\beta x)\ex^{-\alpha x}=\beta^{-\nu}
(\alpha^2+\beta^2)^{-\frac{1}{2}}[(\alpha^2+\beta^2)^\frac{1}{2}-\alpha]^\nu$
\cite{gradshteyn},
leads to the expression for the correlation time:
$$
\frac{\tau_P}{\tau_L}=
\frac{2}{3}\int_0^\infty\frac{\d x}{x^\frac{4}{3}}\Big(\frac{\bar\gamma x^\frac{1}{3}}
{\bar u_G(x-1)^\frac{1}{2}}\Big)^2\Big[\Big(\Big(\frac{\bar\gamma x^\frac{1}{3}}
{\bar u_G(x-1)^\frac{1}{2}}\Big)^2+1\Big)^\frac{1}{2}\Big)\Big]
\eqno(4.27)
$$
We can obtain limiting expressions for this ratio, when the crossing time $(k_0u_G)^{-1}$ is much 
longer or much shorter than the integral time $\tau_L$:
$$
\frac{\tau_P}{\tau_L}=
\begin{cases}
1+\frac{2}{3}(u_Gk_0\tau_L)^2
\log u_Gk_0\tau_L
\quad
&k_0u_G\tau_L\ll 1 
\\
2^\frac{3}{2}(u_Gk_0\tau_L)^{-1}
&k_0u_G\tau_L\gg 1
\end{cases}
\eqno(4.28)
$$

\vskip 10pt
\noindent{\bf 3. Eulerian correlations}
\vskip 5pt
The limit $\tau_S\to\infty$, corresponding to the case of a particle with infinite inertia,
leads, from Eqn. (4.2), to a particle velocity, which, in the absence of gravity, is 
identically zero. Hence $\u^\smalP(\x,t)=\u(\x,t)$ and the time statistics for the fluid velocity 
seen by the 
particle coincides with the Eulerian turbulent statistics. In this regime, the dimensionless
units introduced for Eqn. (4.13) are not appropriate any more. Redefining $\bar t=\gamma_{k_0}t$,
Eqn. (4.13) takes the form, after writing $\exp(-t/\tau_S)\simeq 1-t/\tau_S$:
$$
\langle u_1(0,0)u_1(0,t)\rangle
=\frac{u_T^2}{6}
\int_1^\infty\d x\  x^{-\frac{4}{3}}
\exp\Big[-\bar t x^\frac{1}{3}
-\frac{(x-1)}{2\rho^2}\Big(\frac{3\bar t}{2}-1
+\int_1^\infty\d y\ y^{-2}\exp(-\bar ty^\frac{1}{3})\Big)\Big]
\eqno(4.29)
$$
We start by calculating the Eulerian correlation time 
$$
\tau_E=
u_T^{-2}
\int\d t\langle\u(\x,t)\cdot\u(\x,0)\rangle
\eqno(4.30)
$$
Contrary to  Eqn. (4.13), it is the linear in $t$, $O(\rho^{-2})$ term in Eqn. (4.29), which, 
at fixed long enough $t$, dominates for $x\to\infty$. The same reasons leading to expand Eqn. 
(4.21), suggest that we must now expand:
$$
\exp\Big[\frac{(x-1)}{2\rho^2}\Big(1-
\int_1^\infty\d y\ y^{-2}\exp(-\bar ty^\frac{1}{3})\Big)\Big]
\eqno(4.31)
$$
Instead of Eqn. (4.22), we find:
$$
\int_0^\infty\d t\langle u_1(0,0)u_1(0,t)\rangle
=\frac{u_T^2}{6}
\int_1^\infty\d x\  x^{-\frac{4}{3}}
\Big[\Big(1+\frac{(x-1)^2}{2\rho^2}\Big)\Big(x^\frac{1}{3}+\frac{3(x-1)}{4\rho^2}\Big)^{-1}
$$
$$
-\frac{(x-1)}{2\rho^2}
\int_1^\infty\d y\ y^{-2}\Big(x^\frac{1}{3}+y^\frac{2}{3}+\frac{3(x-1)}{4\rho^2}\Big)^{-1}
\Big]
\eqno(4.32)
$$
All the terms involving factors $\rho^{-2}$ lead, after integration, to an $\O(\rho^{-2})$ result,
except one which leads to a $\O(\rho^{-2}\log\rho)$ term;  the
integral in Eqn. (4.32) will read, to leading order in $\rho$:
$$
\int_1^\infty\d x\ \Big[x^{-\frac{5}{3}}
-\rho^{-2}x^{-1}\Big(1+\rho^{-2}x^\frac{2}{3}\Big)^{-1}\Big]+\O(\rho^{-2})
\simeq \frac{3}{2}-\frac{3\log\rho}{\rho^2}
\eqno(4.33)
$$
We obtain then the result for the Eulerian correlation time:
$$
\frac{\tau_E}{\tau_L}=1-\frac{2\log\rho}{\rho^2}
\eqno(4.34)
$$
which is shorter than $\tau_L$, as expected from the fact that the velocity field statistics
is defined along fluid trajectories, and, sampling at fixed space position should lead to 
an increase in the rate of decorrelation.
Comparing Eqns. (4.28) and (4.36), we see therefore that there is a transition from a correlation
time longer than $\tau_L$ for light particles, to a shorter one for heavy particles. The origin
of this lies in the opposite orderings $\tau_S\lesssim t$ and $\tau_S\gg t$,
on which the Taylor expansions of Eqns. (4.21) and (4.31) are based.
$[$More precisely, for $\tau_S>\rho\tau_L$, we have $k_0S_l>1$ and the
saddle point in Eqn. (4.22) disappears$]$.

As a last exercise, it is possible to calculate the sweep produced decay in an Eulerian
two-point two-time structure function in the form:
$$
S_{rr}(r,t)=\langle [u_r(\r,t)-u_r(0,t)] [u_r(\r,0)-u_r(0,0)]\rangle 
\eqno(4.35)
$$
From the discussion leading from Eqn. (3.14) to Eqn. (3.17), one finds that the structure
function in Eqn. (4.35) is obtained by inserting a factor $2[1-\J_0(rx)-\J_2(rx)]$
in the integrand of Eqn. (4.29). If one considers shorter
time and space scales $k_0r\ll 1$, $t\ll\tau_L$, the leading cause of correlation
decay is sweep, and the $\bar tx^\frac{1}{3}$ in the integrand of Eqn, (4.29) can 
be disregarded. Again because of shortness of $t/\tau_L$, one can Taylor expand 
$\exp(-\bar ty^\frac{1}{3})$ in the same equation and the final result is:
$$
S_{rr}(r,t)
\simeq 2\Ckol\flux^\frac{2}{3}r^\frac{2}{3}\int_0^\infty\d x\
x^{-\frac{5}{3}}[1-\J_0(x)-\J_2(x)]
\exp\Big[-\frac{u_T^2t^2x^2}{6r^2}\Big]
\eqno(4.36)
$$
The term in the exponent is $\O(t/T_{r^{-1}})^2$, with $T_{r^{-1}}$ the 
sweep time at scale $r$. Hence, if $t\gg T_{r^{-1}}$, it is possible to Taylor
expand the Bessel functions and the result is
$$
S_{rr}(r,t)\sim
S_{rr}(r,0) \int_0^\infty \d x\ x^\frac{1}{3}\exp\Big(-(t/T_{r^{-1}})^2x^2\Big)
\sim S_{rr}(r,0)\Big(\frac{T_{r^{-1}}}{t}\Big)^\frac{4}{3}
\eqno(4.37)
$$
i.e. a power law decay of the structure function for times longer than the
sweep time at that space separation.

\vskip 10pt
\centerline{\bf V. Solid tracers: concentration fluctuations}
\vskip 5pt
Because of inertia, the particle velocity field $\v(\x,t)$, contrary to $\u(\x,t)$,
does not preserve volume. Physical intuition suggests that particles which are denser 
than the fluid,
will tend to concentrate near the instantaneous hyperbolic points of the flow, and
to escape from the elliptic ones \cite{wang92,paradisi01}. For this reason, a distribution 
$\theta(\x,t)$ 
of solid particles, in the absence of external sources, will be characterized by finite amplitude
fluctuations superimposed to a uniform mean concentration field $\bar\theta$. These fluctuations are
expected to have a correlation time of the order of $\tau_S$ and a correlation length 
determined in consequence.  We are going to neglect any effect of gravity and set
from the start $\u_G=0$. We will also limit our analysis to the case in which $\tau_S$ is 
in the turbulent inertial range, i.e. we consider $\tau_S\ll\tau_L$ (more precisely, 
$\tau_S<\rho^{-2}\tau_L$). In this way, all
non-universal effects associated with the large scales of the flow are eliminated from the
problem.

The length $S_i$ is crucial to the two-particle statistics, in that it gives the
scale below which solid particles move ballistically relative to one another. In fact, 
$S_c$ fixes the cross-over scale to ballistic behavior, only for the relative motion
of solid and fluid particles; the resulting picture is given by pairs of particles, separated
by $S_i$, moving ballistically over scale $S_c$.
It is easy to see this: if 
$\Delta_r v$ is the typical relative velocity between two solid particles at 
separation $r$ and $\Delta_ru\sim\Ckol^\frac{1}{2}(\flux r)^\frac{1}{3}$ is the corresponding 
value for the fluid velocity, one will have for $r\ll S$, from Eqn. (4.2): $\Delta_r v\sim
(\tau_S\gamma_{r^{-1}})^{-\frac{1}{2}}\Delta_r u$; exploiting the fact that the characteristic
time of variation for $v$ is $\tau_S$, the condition  $\tau_S\Delta_r v\sim r$, gives then
$r\sim S_i$. 
\vskip 5pt

The concentration correlation $\Theta(\r)=\langle\theta(\r,t)\theta(0,t)\rangle$
is proportional to the equilibrium PDF $P(\r)$ for the separation of a pair of solid particles
advected by $\u(\x,t)$. The separation $\r(t)$
obeys an equation in the form
$\dot\r(t)=\v^\smalP(\x+\r,t)-\v^\smalP(\x,t)$ $[$we use from now on the shorthand
$\r(t) \equiv\delta\z^\smalP(t|\x+\r,0)]$, and, for $r\gg S_i$, the separation process takes
a diffusive nature:
$$
\frac{\d}{\d t}\langle[r_\alpha(t)-r_\alpha(0)][r_\beta(t)-r_\beta(0)]\rangle=
2D_{\alpha\beta}(\r)
\eqno(5.1)
$$
A finite level of concentration fluctuations, in the absence of external sources, is associated
with a finite divergence of the diffusivity tensor:
$\partial_\alpha D_{\alpha\beta}\ne 0$. If this component of the diffusivity tensor is small, 
it is possible to proceed perturbatively:
$D_{\alpha\beta}=D_{\alpha\beta}^\smalze+D_{\alpha\beta}^\smalun$, $P=P^\smalze+P^\smalun$, with
$\partial_\alpha D_{\alpha\beta}^\smalze=0$,
$P^\smalze$ uniform and $P^\smalun(\r)\propto \langle  [\theta(\r,t)-\theta(0,t]^2\rangle$;
the equation for the fluctuation amplitude $P^\smalun(\r)$ would read therefore:
$$
D_{\alpha\beta}^\smalze\partial_\alpha\partial_\beta P^\smalun=
-P^\smalze\partial_\alpha\partial_\beta D_{\alpha\beta}^\smalun
\eqno(5.2)
$$
The procedure to determine $D_{\alpha\beta}$ is similar to the one leading to Eqn. (3.17).
From Eqn.  (4.4) and the relation $\dot\r(t)=\v^\smalP(\x+\r,t)-\v^\smalP(\x,t)$, we obtain:
$$
D_{\alpha\beta}(\r)=\lim_{T\to\infty}\frac{1}{T}\int_0^T\d t_1\int_0^T\d t_2
\int_{-\infty}^{t_1}\frac{\d\tau_1}{\tau_S}\int_{-\infty}^{t_2}\frac{\d\tau_2}{\tau_S}
\exp\Big(-\frac{t_1+t_2-\tau_1-\tau_2}{\tau_S}\Big)S_{\alpha\beta}^\smalP(\r,\tau_1,\tau_2)
\eqno(5.3)
$$
with $S_{\alpha\beta}^\smalP$ the time correlation of velocity differences along solid particle
trajectories:
$$
S_{\alpha\beta}^\smalP(\r,t_1,t_2)=
\langle [u_\alpha^\smalP(\r,t_1)-u_\alpha^\smalP(0,t_1)]
[u_\beta^\smalP(\r,t_2)-u_\beta^\smalP(0,t_2)]\rangle
$$
$$
=2[\langle u_\alpha^\smalP(\r,t_1)u_\beta^\smalP(\r,t_2)\rangle
-\langle u_\alpha^\smalP(\r,t_1)u_\beta^\smalP(0,t_2)\rangle]
\eqno(5.4)
$$
We notice that, if we approximated $S_{\alpha\beta}^\smalP(\r,t_1,t_2)
=S_{\alpha\beta}^\smalL(\r,t_1-t_2)$,
since $\partial_\alpha S_{\alpha\beta}^\smalL(\r,t_1-t_2)=0$, we would obtain from Eqn. (5.3) a
divergenceless $D_{\alpha\beta}(\r)$. We have to take into account therefore the effect of
trajectory separation described in Eqn. (4.7).
Proceeding as in the case of the 1-particle statistics,
we arrive at the following modification of Eqn. (4.10):
$$
\langle u_\alpha^\smalP(0,t_1)u_\beta^\smalP(\r,t_2)\rangle
=
-\int\frac{\d^2k}{(2\pi)^2}\frac{\d^2p}{(2\pi)^2}
\exp(\i\p\cdot\r)\frac{\delta^2Z[{\bf J}]}{\delta J_{\k\alpha}(t_1)\delta J_{\p\beta}(t_2)}
\Big|_{{\bf J}=\p\tilde J_{\r t_1t_2}}
\eqno(5.5)
$$
where
$$
\tilde J_{\s,\r t_1t_2}(\tau)=
\bar J_{t_1}(\tau)-\ex^{\i\s\cdot\r}\bar J_{t_2}(\tau)
\eqno(5.6)
$$
and $Z[{\bf J}]$ and $\bar J_t$ are given in Eqns. (4.10-11).
Carrying out the wavevector and time integrations in the definition of $Z[{\bf J}]$ and using
Eqns. (5.6) and (3.1) leads, after some algebra, to the following expression for the velocity
correlation:
$$
\langle u_\alpha^\smalP(0,t_1)u_\beta^\smalP(\r,t_2)\rangle=
\frac{\Ckol\flux^\frac{2}{3}}{\pi}
\int_0^\infty
\frac{k\d k}
{(k^2+k_0^2)^\frac{4}{3}}
\exp(-\gamma_k|t_1-t_2|)
$$
$$
\times\int_0^{2\pi}\d\phi
\Big[\frac{r_\alpha r_\beta}{r^2}\cos^2\phi+
\Big(\delta_{\alpha\beta}-\frac{r_\alpha r_\beta}{r^2}\Big)\sin^2\phi\Big]
\exp(\i kr\cos\phi)
$$
$$
\times
\exp\Big\{-\Ckol\flux^\frac{2}{3}\tau_S^2k^2
\int_0^\infty \frac{[F(s,t_1,t_2)+G(s,t_1,t_2)(\J_0(sr)+\J_2(sr)\cos 2\phi)]s\d s}
{(k_0^2+s^2)^\frac{4}{3}(1+\gamma_s\tau_S)}
\Big\}
\eqno(5.7)
$$
where  $\phi$ is the angle between $\k$ and $\r$,
$$
F(s,t_1,t_2)=f(s,t_1)+f(s,t_2),
\qquad
G(s,t_1,t_2)=f(s,t_1-t_2)-F(s,t_1,t_2)
\eqno(5.8)
$$
and
$$
f(s,t)=1-\ex^{-|t|/\tau_S}-\frac{\ex^{-\gamma_s|t|}-\ex^{-|t|/\tau_S}}
{1-\gamma_S}
\eqno(5.9)
$$
The effect of trajectory separation is contained in the last line of Eqn. (5.7). We see
that the contribution, which leads to finite divergence of the correlation
$\langle u_\alpha^\smalP(0,t_1)u_\beta^\smalP(\r,t_2)\rangle$, is the $\phi$ dependence of this
factor. The remaining $\phi$ dependence, contained in the second line of this equation,
is simply the factor $\k_\perp\k_\perp\exp(\i\k\cdot\r)$ arising in the Fourier transform
of Eqn. (3.1), and would give by itself zero divergence.

The argument of the exponential in the last line of Eqn. (5.7), for fixed
$\tau_S/\tau_L$, is $O(\rho^{-2})$, so that we may try a Taylor expansion. However,
as it happened with Eqns. (4.21) and (4.31), the resulting integrals in $k$ diverge.
We therefore keep in the exponential the leading
contribution in $k$, which is the time independent piece of its argument,
and expand the remnant, which, to leading
order in $\rho$, gives the following expression:
$$
1-\Ckol\flux^\frac{2}{3}\tau_S^2k^2\cos 2\phi
\exp\Big\{-\int_0^\infty\frac{\Ckol\flux^\frac{2}{3}\tau_S^2k^2s\d s}
{(k_0^2+s^2)^\frac{4}{3}(1+\gamma_s\tau_S)}\Big\}
\int_0^\infty \frac{G(s,t_1,t_2)\J_2(sr)s\d s}
{(k_0^2+s^2)^\frac{4}{3}(1+\gamma_s\tau_S)}
$$
$$
=1-\Ckol\flux^\frac{2}{3}\tau_S^2k^2\cos 2\phi
\exp\Big\{-\frac{3\Ckol\flux^\frac{2}{3}\tau_S^2k^2}{2k_0^\frac{2}{3}}
\Big\}
\int_0^\infty \frac{G(s,t_1,t_2)\J_2(sr)s\d s}
{(k_0^2+s^2)^\frac{4}{3}(1+\gamma_s\tau_S)}
\eqno(5.10)
$$
plus terms which would lead to a divergence free contribution to
$\langle u_\alpha^\smalP(0,t_1)u_\beta^\smalP(\r,t_2)\rangle$ and would disappear from Eqn. (5.2).
Substituting into Eqn. (5.7) and then back into Eqns. (5.4) and (5.3), we find, after
carrying out the time integrals and
the integral in $\phi$:
$$
D^\smalun_{\alpha\beta}=\frac{8\Ckol^\frac{3}{2}\flux\tau_S^2}{\rho}
\int_0^\infty x^{-\frac{1}{3}}\d x
\exp\Big\{-\frac{3S_l^2x^2}{2r^2}\Big\}
\int_0^\infty\d y\ y^{-\frac{5}{3}}\J_2(y)\Big[1-\frac{1}{2(1+(x/y)^\frac{2}{3})}\Big]
$$
$$
\times\Big[\delta_{\alpha\beta}(\frac{1}{2}\J_0(x)-\J_2(x)+\frac{1}{2}\J_4(x))
-\frac{r_\alpha r_\beta}{r^2}\J_4(x))\Big]
\eqno(5.11)
$$
and it is possible to see that $D^\smalze_{\alpha\beta}$ is given by the same expression 
valid for a fluid parcel, i.e. by Eqn. (3.17):
$$
D^\smalze_{\alpha\beta}(\r)=
\frac{4\alpha_\smalsev\Ckol^\frac{1}{2}\flux^\frac{1}{3}}{\rho}r^\frac{4}{3}
\Big[\frac{r_\alpha r_\beta}{r^2}+\frac{7}{3}\Big(\delta_{\alpha\beta}
-\frac{r_\alpha r_\beta}{r^2}\Big)\Big]
\eqno(5.12)
$$
The physical content of the expansion leading to Eqns. (5.11-12) can be clarified, noticing
that, in a way perfectly analogous to Eqns. (4.10-11), the generating functional $Z[{\bf J}]$
entering Eqn. (5.5) can be written as:
$$
\Big\langle\exp\Big(\i{\bf k}\cdot\int\d t(\u^\smalL(0,t)\bar J_{t_1}(t)
-\u^\smalL(\r,t)\bar J_{t_2}(t))\Big)\Big\rangle
$$
$$
\sim\exp\Big\{-\frac{{\bf k}{\bf k}}{2}:\int\d t\d t'\,[{\bf U}(0,t-t')
(\bar J_{t_2}(t)\bar J_{t_2}(t')+\bar J_{t_1}(t)\bar J_{t_1}(t'))
+2{\bf U}(\r,t-t')J_{t_1}(t)\bar J_{t_1}(t')]\Big\}
$$
The argument in the exponential is in the form $k^2U(\r,0)\tau_S^2\sim k^2(S_l^2+
\Ckol(\flux r)^\frac{2}{3}\tau_S^2)$, where the term involving $S_l$ gives the one-particle 
contribution 
to trajectory separation, while the remnant is the two-particle correction coming from $r>0$.
Substituting into the definition of $D_{\alpha\beta}$ gives then, using $r\ll S_l$:
$$
D\sim\int k\d k U_k\gamma_k^{-1}\ex^{-k^2S_l^2}
\Big[1-\exp\Big(-k^2\Ckol(\flux r)^\frac{2}{3}\tau_S^2
-\i{\bf k}\cdot\r\Big)\Big]
$$
$$
\sim \int k\d k U_k\gamma_k^{-1}(1-\ex^{-\i{\bf k}\cdot\r})
+\int k^3\d k U_k\gamma_k^{-1}\Ckol(\flux r)^\frac{2}{3}\tau_S^2\ex^{-S_l^2k^2}
$$
which, using Eqn. (4.9), is 
$\sim \frac{\Ckol^\frac{1}{2}\flux^\frac{1}{3}}{\rho}[r^\frac{4}{3}
+\rho^2S_i^\frac{4}{3}(\frac{r}{S_l})^\frac{2}{3}]$. Thus, the origin of the $D^\smalun$ in
the two-particle contribution to trajectory separation is confirmed, together with its being
generated at the $k_0$-dependent scale $S_l$. 
This is of course confirmed by direct analysis of Eqn. (5.11); the integral is dominated by 
$k=x/r\sim S_l^{-1}$ and we obtain, for $r>S_i$:
$$
\frac{D^\smalun}{D^\smalze}\sim\frac{\rho^2 S_i^\frac{4}{3}}{(rS_l)^\frac{2}{3}}
<\rho^\frac{2}{3}\Big(\frac{\tau_S}{\tau_L}\Big)^\frac{1}{3}
\eqno(5.13)
$$
Thus, for $\tau_S<\rho^{-2}\tau_L$, $D^\smalun<D^\smalze$ and Eqn. (5.3) applies.

Using the relation: $\int_0^\infty\d x x^{-\frac{1}{3}}\ex^{-\alpha x^2}\J_4(x)=
\frac{\Gamma(7/3)}{2^5\alpha^{7/3}\Gamma(5)}{\rm M}(7/3,5,-1/(4\alpha))$, with 
${\rm M}(a,b,x)$ the confluent hypergeometric function \cite{abramowitz}, we obtain in general, from Eqn. (5.11):
$$
D^\smalun_{\alpha\beta}=
4\rho\alpha_\frac{7}{3}\Ckol^\frac{1}{2}\flux^\frac{1}{3} 
S_i^\frac{4}{3}
\Big(\frac{r_\alpha r_\beta}{r^2} \tilde D(r/S_l)+\delta_{\alpha\beta}\hat D(r/S_l)\Big),
\eqno(5.14)
$$
where we can write:
$$
\tilde D(r/S_l)=-\frac{c\beta\, 2^\frac{2}{3}\Gamma(7/3)}{\alpha_\frac{7}{3}\Gamma(5)}
\Big(\frac{r^2}{6S_l^2}\Big)^\frac{7}{3}
{\rm M}(\frac{7}{3},5,-\frac{r^2}{6S_l^2});
\qquad
\beta=\int_0^\infty
\d y\ y^{-\frac{5}{3}}\J_2(y),
\eqno(5.15)
$$
with $c\simeq 1$ for $r\gg S_l$ and $c\simeq\frac{1}{2}$ for $r\ll S_l$;
$\hat D$ will be shown not to contribute to the concentration correlations.

We can now calculate the probability $P^\smalun(r)$. Substituting Eqns. (5.13-15) 
into Eqn. (5.2), after a few manipulations, leads to:
$$
\partial_{\bar r}\bar r^\frac{7}{3}\partial_{\bar r}P^\smalun=-
\rho^2 
\bar rP^\smalze
\Big(\frac{S_i}{S_l}\Big)^\frac{4}{3}
\Big(\partial_{\bar r}^2(\tilde D(\bar\r)+\hat D(\bar r))
+\frac{1}{\bar r}\partial_{\bar r}(2\tilde D(\bar r)+\hat D(\bar r))\Big)
\eqno(5.16)
$$
where $\bar r=r/S_l$.
Hence, for $S_i\ll r\ll S_l$: 
$$
P^\smalun(r)= \rho^2 P^\smalze
\Big(\frac{S_i}{S_l}\Big)^\frac{4}{3}
\int_{r/S_l}^\infty\d y y^{-\frac{7}{3}}
\Big[\tilde D(\infty)-y\partial_y(\tilde D(y)+\hat D(y))-\tilde D(y)\Big]\simeq
$$
$$
\simeq\frac{3}{4}
\rho^2P^\smalze\tilde D(\infty)\Big(\frac{S_i}{r}\Big)^\frac{4}{3}
\eqno(5.17)
$$
Using Eqn. (5.15) and the limiting form for the confluent hypergeometric function 
${\rm M}(a,b,-z)=\frac{\Gamma(b)}{\Gamma(b-a)}z^{-a}(1+O(z^{-1}))$ \cite{abramowitz}, we get the 
final result:
$$
\Theta(r)=\bar\theta^2\Big(1+
\bar\beta\rho^2\Big(\frac{S_i}{r}\Big)^\frac{4}{3}\Big)
\eqno(5.18)
$$
where
$$
\bar\beta=\frac{3\beta\Gamma(7/3)}{2^\frac{2}{3}\alpha_\frac{7}{3}\Gamma(8/3)}\simeq 2.14
\eqno(5.19)
$$
In conclusion, we have a range of separations $S_i\ll r\ll S_l$, in which the
fluctuation correlation grows with a power $-\frac{4}{3}$, to reach
amplitude $\sim \rho^2$ at $r\sim S_i$.

The picture which arises is one of concentration fluctuations produced at scale
$S_l$, by compressibility of the solid particle flow, and then transported to small scales
and amplified by the incompressible part of the flow. The process is different from that
of a passive scalar forced at large scale, 
due to the derivatives in the source term
$[$and in fact the scaling exponent is different; compare with Eqn. (3.36)$]$.
This source term is basically $\nabla^2D^\smalun(r)$, with $D^\smalun(r)$ saturating at
a constant for $r\gg S_l$ and going
to zero in the opposite limit. From here, the $r^{-\frac{4}{3}}$
scaling of Eqn. (5.18) arises by dimensional analysis. 

What happens when $r\ll S_i$? 
At such short distances, the separation process is ballistic and
we cannot use a diffusive approximation anymore. In \cite{balkovsky01}, it is suggested
that the correlation build-up should stop only because of discreteness effects or because
of the Brownian motion of the solid particle. Actually, extrapolating the results of the
present paper to the real turbulence regime $\rho=O(1)$, there is good reason to think that, for
$\tau_S>\tau_\eta$, 
this build-up could stop much earlier, and precisely at $r\sim S_i$, which, for $\rho=O(1)$, 
coincides with the size of vortices with eddy turnover time $\tau_S$.

At separations below $S_i$, Eqn. (5.18) ceases to be valid, and full analysis 
of the distribution $P(\r,\Delta_r\v)$ is needed.
A singularity of $P(r)$ at $r=0$ would require focusing of $\Delta_r\v$ along $\r$ for $r\ll S_i$;
the mechanism is sketched in Fig. 1.
\begin{figure}[hbpt]\centering
\centerline{
\psfig{figure=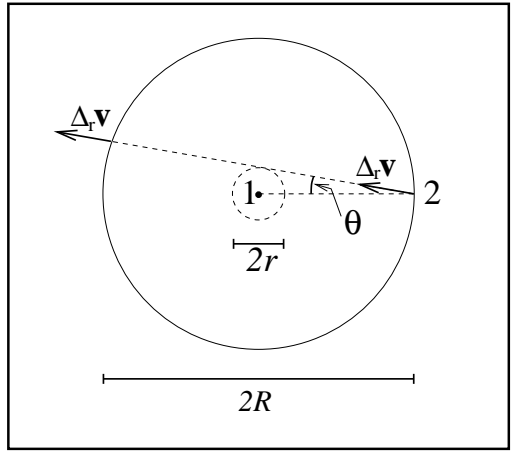,height=6.cm,angle=0.}
}
\caption{Sketch of the behavior of particle relative velocities inside a domain of size
$R<S_i$. Particle 2 moves with respect to particle 1 at constant velocity.
In order for the particle density to diverge as $r\to 0$, it is necessary that the distribution
of velocities be peaked too at $\theta=0$. 
}
\end{figure}
This means that $P(\r,\Delta_r\v)$ itself should develop, as
$r\to 0$, a singularity at $\theta$, where $\theta$ is the angle between $\Delta_r\v$ and $\r$.
The necessary trajectory focusing can be produced only by the compressible part of $\v$.
However, for $r<S_i$, the production term for the compressible part of $\v$ 
can be estimated directly from the second term in Eqn. (5.10), 
to be $O(\rho^{-2})$ relative to the rest, and is able to act only
for a time $\tau_S$ in the ballistic region. Hence, $P(\r,\bar\v)$ must be singular before this
region is reached. On the other hand, for $r>S_i$, where the diffusive approximation works,  the 
asymmetry of $P(\r,\bar\v)$, associated with compressibility of the flow, can be estimated from 
$\frac{\langle\bar v_\alpha\bar v_\beta\rangle^\smalun}
{\langle\bar v_\alpha\bar v_\beta\rangle^\smalze}
\sim D^\smalun/D^\smalze
$
so that, if $\tau_S<\rho^{-2}\tau_L$, singularities should not be expected in $P(\r,\bar\v)$, for $\theta=0$
and $r\ge S_i$ either. The conclusion is that a plateau for $\Theta(r)$ should be 
present at $r<S_i$.

\vskip 10pt
\centerline{\bf VI. Solid tracers: ergodic properties}
\vskip 5pt
One of the consequences of the compressibility of the velocity field $\v(\x,t)$ is that the
ergodic property is not satisfied anymore: velocity moments calculated along solid particle
trajectories differ from those obtained from spatial averages. As mentioned before, physical
intuition suggests that
solid particles should privilege in their motion certain regions of the fluid with respect
to the others (namely, hyperbolic with respect to elliptic regions). It is difficult,
however, to translate this into a statement on the form of the PDF for the velocity $\u^\smalP$.

We have at our disposal the equations satisfied by the velocity field $\u^\smalP$. It is
possible therefore to calculate its moments and to reconstruct its PDF. We consider the
case of zero gravity $\u_G=0$ and $\tau_S/\tau_L$ small. As in the analysis of the concentration
fluctuations, all non-universal effects associated with the large scales of the flow are thus
eliminated from the problem.  From definition
of $\u^\smalP$ and Eqns. (4.1-2), we obtain the following set of equations, valid to
lowest order in $\rho^{-1}$:
$$
\begin{cases}
(\partial_t+\tilde\u(\x,t)\cdot\nabla)\u^\smalP(\x,t)+
\int\d y^2\gamma(\x-y)\u^\smalP(\y,t)
=\int\d^2yh(\x-\y){\boldsymbol\xi}(\y,t)
\\
\tilde\u(\x,t)\equiv\u^\smalP(\x,t)-\v^\smalP(\x,t)=\int_{-\infty}^t\d\tau\exp(-\frac{t-\tau}{\tau_S})
\dot\u^\smalP(\x,t)
\end{cases}
\eqno(6.1)
$$
which differs from the analogous equation for $\u^\smalL$ because of the non volume-preserving
advection term
$\tilde\u\cdot\nabla\u^\smalP$. From here we can carry on standard field theoretical
perturbation theory, either by the Martin-Siggia-Rose formalism \cite{martin73}, or working
directly with Eqn. (6.1). The building blocks of the diagrammatic expansion are shown in Fig. 2,
and are the propagator $G_{\k\alpha\beta}$:
$$
G_{\k\alpha\beta}u^\smalP_{\k\beta}(t)
=\int_{-\infty}^t\d\tau\ \exp(-\gamma_k(t-\tau))\frac{k_\alpha k_\beta}{k^2}
u^\smalP_{\k\beta}(\tau)
\eqno(6.2)
$$
the correlator $U^\smalP_{\k \alpha\beta}(t)$:
$$
U^\smalP_{\k \alpha\beta}(t)=
\frac{k^\perp_\alpha k^\perp_\beta}{k^2}\frac{4\pi\Ckol\flux^\frac{2}{3}}{(k^2+k_0^2)^\frac{4}{3}}
\exp(-\gamma_k|t|)
\eqno(6.3)
$$
and the vertex $\Gamma_{\k\alpha\beta\gamma}$:
$$
\Gamma_{\k\alpha\beta\gamma}u^\smalP_{\p\beta}u^\smalP_{\s\gamma}(t)=
\i\lambda s_\beta\delta_{\alpha\gamma}\delta(\k+\p+\s) \int_{-\infty}^t\d\tau\
\exp(-\frac{t-\tau}{\tau_S})
u^\smalP_{\s\gamma}(\tau)\partial_\tau u^\smalP_{\p\beta}(\tau)
\eqno(6.4)
$$
\begin{figure}[hbpt]\centering
\centerline{
\psfig{figure=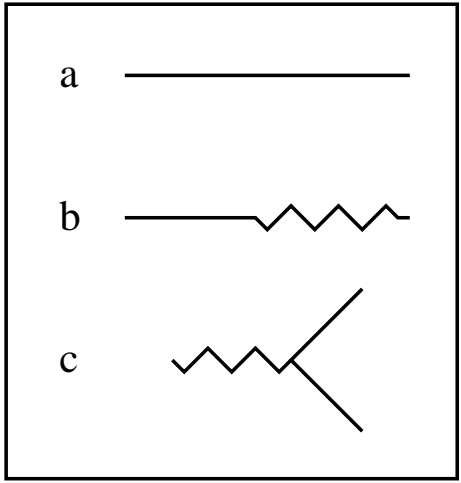,height=6.cm,angle=-90.}
}
\caption{Feynman diagrams for the propagator $G_{\k\alpha\beta}$ (a), for the correlator
$U^\smalP_{\k \alpha\beta}$ (b) and the vertex $\Gamma_{\k \alpha\beta\gamma}$
(c).
}
\end{figure}
where the coefficient $\lambda=1$ is introduced, as in Eqn. (2.15), only for the purpose
of book-keeping.
To lowest order in $\lambda$, the
correlations for the fields $\u^\smalP(\x,t)$ and $\u^\smalL(\x,t)$
are trivially equal. To higher orders, differences arise, which would not lead,
if $\nabla\cdot\tilde\u=0$, to differences between the one-point PDF's for
$\u^\smalP$ and $\u^\smalL$ (see also \cite{tennekes}). In our case, this is
not so, and the difference between
the moments of the two PDF's can be calculated in perturbation theory; to $\O(\lambda^n)$:
$$
\langle (u^\smalP)^m\rangle^\smaln=
\int\frac{\d^2k_1 }{(2\pi)^2}...\frac{\d^2k_m }{(2\pi)^2}
\langle u_{\k_1}(t)...u_{\k_m}(t)\rangle^\smaln
\eqno(6.5)
$$
where $\langle u_{\k_1}(t)...u_{\k_m}(t)\rangle^\smaln$ is the sum of the Feynman
diagrams with
$m$ outgoing velocity lines and $n$ vertices. Because of symmetry under space reflection,
the lowest order contributions are $\O(\lambda^2)$; the corresponding diagrams are
shown in Fig. 3 and lead to corrections to the velocity second and fourth moments.
\begin{figure}[hbpt]\centering
\centerline{
\psfig{figure=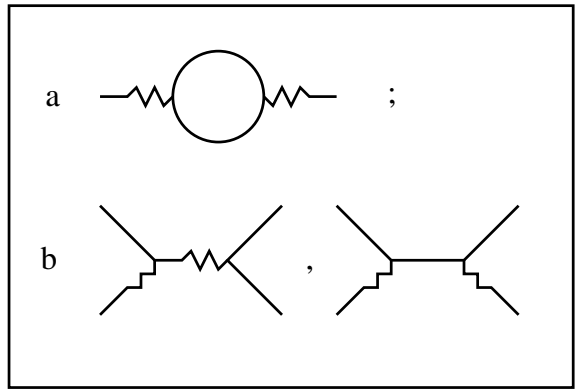,height=7.5cm,angle=-90.}
}
\caption{Feynman diagrams providing the lowest order correction to
$\langle (u^\smalP)^2\rangle$ (a) and $\langle (u^\smalP)^4\rangle$ (b).
}
\end{figure}
In order to check for the presence of divergences in loop diagrams, we carry
on power counting on Eqn. (6.1). Rescaling space and time as in Eqn. (2.16), we find
$[\lambda]=0$, implying the possibility of logarithmic divergences. Now, a perturbation
expansion in $\lambda$ of Eqn. (6.1) ceases to be sensible at scales below the length
$S_i$ defined in Eqn. (4.9). From 
Eqn. (4.13), the effective decay rate
for the field $\u^\smalP$ appears to be:
$$
\gamma^\smalP_k=
\begin{cases}
\gamma_k \quad & kS_i\ll 1
\\
u_Sk\quad & kS_i\gg 1
\end{cases}
\eqno(6.6)
$$
and this expression should be substituted for $\gamma$ in Eqns. (6.1-3).
For $kS_i\gg 1$, $\gammaP_k$ is just the inverse of the crossing time of an eddy
of size $k^{-1}$. In this large
$k$ range, the appropriate scaling for the frequency should be, instead of the one
provided by Eqn. (2.17), which lead to $[\lambda]=0$, the following one:
$$
[t]=1,
\qquad
[\lambda]=-[u]=-\frac{1}{2}
\eqno(6.7)
$$
The change of scaling in $\gammaP$ is therefore sufficient to regularize the
divergent diagrams, providing an effective ultraviolet
cutoff
at $k=S_i^{-1}$.

We calculate explicitly the loop diagram in Fig. 3, and the corresponding
correction to $\langle (u^\smalP)^2\rangle$:
$$
\langle (u^\smalP)^2\rangle^\smaldu=
\int\frac{\d^2p}{(2\pi)^2}\frac{\d^2s}{(2\pi)^2}
\int_{-\infty}^0\d t_1\int_{-\infty}^0\d t_2
\int_{-\infty}^{t_1}\d\tau_1\int_{-\infty}^{t_2}\d\tau_2
$$
$$
\times\exp\Big(-\gammaP_k(t_1+t_2)
-\frac{t_1+t_2-\tau_1-\tau_2}{\tau_S}\Big)
\delta_{\alpha\beta}
\partial_{\tau_1}\partial_{\tau_2}s_\gamma s_\delta
[U^\smalP_{\p\gamma\delta}(\tau_1-\tau_2)U^\smalP_{\s\alpha\beta}(t_1-t_2)
$$
$$
+U^\smalP_{\p\gamma\beta}(\tau_1-t_2)U^\smalP_{\s\alpha\delta}(t_1-\tau_2)
+U^\smalP_{\p\alpha\gamma}(t_1-\tau_2)U^\smalP_{\s\alpha\beta}(\tau_1-t_2)
+U^\smalP_{\p\alpha\beta}(t_1-t_2)U^\smalP_{\s\gamma\delta}(\tau_1-\tau_2)]
$$
where $\k=-\p-\s$. The time integrations can be carried out at once and, after some
algebra, we reach the following result:
$$
\langle (u^\smalP)^2\rangle^\smaldu=
2\int\frac{\d^2p}{(2\pi)^2}\frac{\d^2s}{(2\pi)^2}
\frac{\UP_p\UP_s\gammaP_p(\p_\perp\cdot\s)^2}
{(\gammaP_k+\gammaP_p+\gammaP_s)(\gammaP_k+\gammaP_s+\tau_S^{-1})(\gammaP_p+\tau_S^{-1})
\gammaP_kp^2}
$$
$$
\times\Big[-\gammaP_p\tau_S(\gammaP_k+\gammaP_p+\gammaP_s+\tau_S^{-1})
+\frac{(\p\cdot\s)}{s^2}\frac{\gammaP_s}{\gammaP_s+\tau_S^{-1}}
(\gammaP_k+\gammaP_s-\tau_S^{-1})\Big]
\eqno(6.8)
$$
where $U_k=U_{\k \alpha\alpha}(0)$. As predicted in the discussion leading to Eqns. (6.6-7), substituting
$\gammaP\to\gamma$ would lead to a logarithmically divergent integral. 
Comparing with Eqn. (4.9), we see that this integral receives contribution from wavevectors in
the range $[S^{-1},S_i^{-1}]$, i.e. from those eddies fast enough for the particles to be
unable to respond to their velocity field, but still sufficiently slow for trajectory separation
to be considered a perturbation. To find the leading
behavior in $S_i^{-1}$, the integral can be rewritten, after the change of variables
$y=(\gamma_{(ps)^\frac{1}{2}}\tau_S)^{-1}$,
$z=p/q$, in the form:
$$
\langle (u^\smalP)^2\rangle^\smaldu=
\frac{3u_S^2}{16\pi^3\rho^2}\int_0^{2\pi}\d\phi\int_0^\infty\frac{\d z}{z}\int_{\rho^{-2}}^\infty
\frac{\d y}{y}\frac{\sin^2\phi}
{(\bar p^\frac{2}{3}+\bar s^\frac{2}{3}+\bar k^\frac{2}{3})
(\bar k^\frac{2}{3}+\bar s^\frac{2}{3}+y)
(\bar p^\frac{2}{3}+y)\bar k^\frac{2}{3}\bar p^\frac{2}{3}}
$$
$$
\times\Big[-(\bar p^\frac{2}{3}+\bar s^\frac{2}{3}+\bar k^\frac{2}{3}+y)
+\frac{y\cos\phi}{\bar s^\frac{2}{3}+y}(\bar s^\frac{2}{3}+\bar k^\frac{2}{3}-y)\Big]
+\O(\rho^{-2})
\eqno(6.9)
$$
where $\bar\k=(ps)^{-\frac{1}{2}}\k$, $\bar\p=(ps)^{-\frac{1}{2}}\p$,
$\bar\s=(ps)^{-\frac{1}{2}}\s$, $\cos\phi=\bar\p\cdot\bar\s$. We obtain then the final
result:
$$
\langle (u^\smalP)^2\rangle^\smaldu=
\frac{\bar\eta\log\rho}{\rho^2}u_S^2
\eqno(6.10)
$$
where:
$$
\bar\eta=\frac{3}{8\pi^3}\int_0^{2\pi}\d\phi\int_0^\infty\frac{\d z}{z}
\frac{z^\frac{1}{3}\sin^2\phi}{(z+z^{-1}+2\cos\phi)^\frac{1}{3}}
\Big[-\frac{1}{(z+z^{-1}+2\cos\phi)^\frac{1}{3}+z^\frac{1}{3}}
$$
$$
+\frac{\cos\phi}{z^\frac{1}{3}+z^{-\frac{1}{3}}+(z+z^{-1}+2\cos\phi)^\frac{1}{3}}\Big]
\simeq -0.32
\eqno(6.11)
$$
is evaluated by numerical integration.
The correction to the velocity amplitude is negative.
In the presence of inertia, solid tracers prefer therefore to lie in regions of the
flow where the turbulent velocity is smaller.

Extrapolating Eqn. (6.10) to $\rho=O(1)$ suggests that $\langle (u^\smalP)^2\rangle-
u_T^2\sim u_S^2$. We can have some idea of what we should expect for dominant gravity
$u_S<u_G$ from dimensional analysis of Eqn. (6.8). In this case $\gamma_k\to u_Gk$,
the inverse sweep time due to the particle fall, and we would find
$\langle (u^\smalP)^2\rangle^\smaldu\sim \frac{k^4 U_k^2}{u_G^2}$ with $k^{-1}\sim u_G\tau_S$
giving the transition to the small scales for which the sweep time is shorter than $\tau_S$,
and to which the particles are unable to respond. From here we find
$\langle (u^\smalP)^2\rangle-
u_T^2\sim (u_S/u_G)^\frac{2}{3}u_S^2$ and we see that gravity reduces the amount of non-ergodicity
of the solid particle flow.

\vskip 10pt
\centerline{\bf VII. Conclusions}
\vskip 5pt
Consideration of a finite correlation time in the transport by a random velocity
field has allowed analysis of a series of issues. We summarize the main results:
\newcounter{rom}
\begin{list}
{\roman{rom}}{\usecounter{rom}\setlength{\rightmargin}{\leftmargin}}
\item The self-diffusion of a fluid parcel obeys linear scaling in the inertial range (as it
should) with a universal constant $C_0=\Ckol^\frac{3}{2}
\frac{\hat\rho\rho}{\hat\rho-\rho}\log\hat\rho/\rho$ $[$see Eqns. (3.7-3.10)$]$, which is sensitive
both to the ratio of the eddy turn-over and life time, and to the rate of eddy velocity 
decorrelation at times much shorter than the eddy lifetime.
A quadratic maximum (at least), at time separation equal to zero, is necessary for $C_0$ to 
remain
finite. (An exponential time correlation, for instance, would not satisfy this condition).
This sensitivity on the short time behavior of time correlation was not observed in any
of the other transport processes considered in the present paper.
\item The relative diffusion of a pair of fluid parcels, exhibits (again as it should) 
Richardson and normal diffusion behavior, respectively, for coordinates and velocities. The 
PDF for relative separation is a stretched exponential with exponent $\frac{2}{3}$ $[$see Eqn. 
(3.21)$]$ and it is possible to express the universal constants $c$ and $\tilde c$ entering 
respectively coordinate and velocity dispersion, in terms of the parameter $\rho$. Precisely:
$c\simeq\frac{0.748\Ckol^\frac{3}{2}}{\rho^3}$ and
$\tilde c\simeq\frac{3.037\Ckol^\frac{3}{2}}{\rho}$ $[$see Eqns. (3.23) 
and (3.26)$]$. For the Batchelor constant, we obtain instead $[$see Eqn. (3.39)$]$:
$B\simeq 11.32\rho$.
\item The correlation time $\tau_P$ for the fluid velocity sampled by a solid particle has a 
behavior consistent with previous analysis
neglecting the structure of the turbulent inertial range \cite{csanady63,reeks77}.
Values of $\tau_P/\tau_L$,
above unity are found for dominant inertia and
$\tau_S\lesssim\tau_L$, with $\tau_P/\tau_L-1=O((\frac{\tau_S}{\rho\tau_L})^\frac{4}{3})$
$[$see Eqn. (4.25)$]$. 
On the contrary, in the case
of dominant gravity, $\tau_P/\tau_L<1$ irrespective of the value of the ratio $u_T/u_G$ 
between the turbulent and the fall velocity; specifically $[$see Eqn. (4.28)$]$, we find 
$\tau_P/\tau_L-1=O((u_Gk_0\tau_L)^2)$ for $u_G\ll u_L$, and $\tau_P/\tau_L=O((u_Gk_0\tau_L)^{-1})$
in the opposite case. For short times, the expected sub-linear behavior for the fluid
velocity along a solid particle trajectory is found: 
$\langle |u^\smalP(x,t)-u^\smalP(x,0)|^2\rangle\sim(\flux u_At)^\frac{2}{3}$, with $A=G,S$
depending on whether gravity or inertia dominates $[$see Eqns. (4.16) and (4.19)$]$.
\item The Eulerian correlation time $\tau_E$ (and by continuity, therefore, also 
$\tau_P$, in the regime $\tau_S\gg\tau_L$) is shorter than its Lagrangian
counterpart, with $\tau_E/\tau_L=1-2\rho^{-2}\log\rho$ $[$see Eqn. (4.34)$]$. Sweep produces a power law decay
of correlations between velocity increments in the form
$S_{rr}(r,t)=\langle [u_r(\r,t)-u_r(0,t)] [u_r(\r,0)-u_r(0,0)]\rangle$. More precisely, for
time separations longer than the sweep time $T_{r^{-1}}$:
$S_{rr}(r,t)\sim S_{rr}(r,0)(T_{r^{-1}}/t)^\frac{4}{3}$ $[$see Eqn. (4.37)$]$.
\item In the absence of gravity, and for $\rho^2\tau_\eta\ll\tau_S\ll\rho^{-2}\tau_L$, the spectrum of 
concentration correlation induced by turbulence in a solid particle suspension, is 
universal and has power law behavior for separations above the size $S_i$ of an 
eddy which is crossed by a typical solid particle in a time equal to its lifetime. More precisely:
$\bar\theta^{-2}\langle\theta(r)\theta(0)\rangle-1\simeq\rho^2
(S_i/r)^\frac{4}{3}$ $[$see Eqn. (5.18)$]$.
\item The solid particle flow is non-ergodic, with a difference between the fluid velocity sampled
along a solid trajectory and the corresponding Eulerian average:
$\langle (u^\smalP)^2\rangle-u_T^2=-\frac{0.32\log\rho}{\rho^2}u_S^2$ $[$see Eqns.
(6.10) and (6.11)$]$. Dimensional reasoning for $\rho=O(1)$ suggests that gravity should reduce 
this effect from $\langle (u^\smalP)^2\rangle-u_T^2\sim u_S^2$ to 
$\langle (u^\smalP)^2\rangle-u_T^2\sim (u_S/u_G)^\frac{2}{3}u_S^2$.
\end{list}

Analysis of some of these problems, actually did not exploit the finite correlation time of the
velocity field produced through Eqn. (2.7). In particular, 
the process of fluid parcel relative dispersion was considered to the same order
in $\rho$ as in the Kraichnan model and, to this level, no information on the Lagrangian 
statistics was necessary. Finiteness of the correlation time had
the only purpose to allow a meaningful definition of quantities such as $\Ckol$ and
$\flux$.

In the case of the self-diffusion properties of fluid and solid particles, 
a finite correlation time and inclusion of the 
Lagrangian nature of time correlation was necessary from the start. Nonetheless,
the only point in which analysis of the random velocity field could not be avoided,
was to determine the dimensionless constant $C_0$ \cite{thomson87,sawford91}; 
the diffusion exponents in the various cases were already available by dimensional 
reasoning.

Evaluation of the correlation time $\tau_P$ and analysis of concentration fluctuations
and non-ergodicity of particle trajectories (points iv-vi), instead, rested heavily 
on the fact that the correlation time was finite and on knowledge of the actual form of the 
random velocity field time correlation. The analysis confirmed the role 
of eddies with lifetime $\tau_S$, already pointed out in \cite{olla01}. 

Some comments are due on these last issues.
As regards correlation times, they depend in general on non-universal aspects of the velocity
statistics, and, in the present case, on the assumption that also the large scale 
statistics is defined along Lagrangian trajectories. In consequence of this, the 
Eulerian time of the flow resulted shorter than the Lagrangian correlation time.
(Following \cite{kraichnan64a}, the Eulerian correlation feels, at the same time, the decorrelation 
from relative motion of the fluid, and the effect of eddy decay). For $\tau_S\ll\tau_L$, 
the standard picture of inertia and gravity leading, respectively, to increase
and decrease of the correlation time, however, was confirmed.

As regards concentration fluctuations, previous treatments of this problem, either were limited 
to the case of particles with Stokes time shorter than the Kolmogorov time of the flow 
\cite{balkovsky01}, or neglected turbulent small scale structures altogether \cite{elperin00}.
This was due to the 
difficulty in analyzing trajectory crossing effects on inertial range scales, associated with
the need for a proper treatment of the Lagrangian time statistics. The fully kinetic treatment
adopted here, in which the relative motion of individual solid particles is fully taken 
into account, in contrast with the fluid equation approach used in \cite{balkovsky01}, together
with the large $\rho$ limit, is 
what allows treatment of the problem. 

It should be mentioned that solid particle concentration 
fluctuations may be important in the process of rain formation.
It is known that the settling rate of a suspension is enhanced in the presence of clumping
of the heavy particles \cite{wang93a}, and turbulence induced concentration fluctuations appear
to be one of the important actors in the process \cite{aliseda00}. Inclusion of the effect
of gravity, on the same lines of the analysis carried on in section IV would therefore be
necessary.

As regards non-ergodicity of the solid particle flow, it should be mentioned that this 
is a problem 
one has to deal with, before trying to extend standard Lagrangian transport models
(in particular the well mixedness hypothesis on which they are based \cite{thomson87}) 
to the case of solid particles. 

An important aspect that must be stressed, in the calculation of both $\tau_P$ and the 
concentration correlation spectrum, is the role played by the localization length $S_l$.
This length ceases to have a physical meaning for finite $\rho$, nonetheless, it fixes,
in perturbation theory, the scale at which both fluctuations and the difference 
$\tau_P-\tau_L$ are generated. Notice that, in the case of concentration fluctuations, 
this occurs in spite of the fact that the concentration correlations are peaked at the 
inertial scale $S_i$.

Another peculiarity of the large $\rho$ expansion is the multiplicity of space scales 
associated with eddies having time or velocity scales related to $\tau_S$ and $\u_S$ 
$[$see Eqns. (4.6) and (4.9)$]$. All of them collapse, for 
$\rho=O(1)$, on the size of a vortex with turnover time equal to $\tau_S$. 
In real high Reynolds number turbulence, this is the saturation length 
expected for concentration fluctuation build-up, when $\tau_S$ is an inertial range
quantity.

\vskip 5pt
The parameters $\rho$ and $\hat\rho$ are central to the extension of the Kraichnan model
to finite correlation times. The situation of reference in real flows is the inverse 
cascade range of two-dimensional turbulence. An estimate of these parameters could be
obtained using the leading $\rho$ expressions provided by Eqns. (3.10), (3.23), (3.26) 
and (3.39), with the values of the constants $C_0$, $c$, $\tilde c$ and $B$ obtained from DNS.  
For instance, assuming $\hat\rho=\infty$, comparison with the results presented in 
\cite{boffetta00} would give $\rho\simeq 2$.

The results of the present paper have been obtained to leading order in $\rho$. To this order, 
no perturbative
effects in the structure of random velocity fields are present, and the correlations for the
Lagrangian velocity $u^\smalL$ obey Eqn. (2.6). The parameters entering these correlations
must nonetheless be considered as renormalized quantities in a renormalized statistical
field theory. No claim on the nature of these renormalizations is made, apart that, to lowest 
order, marginality of interactions suggests that correction to scaling be only logarithmic.

To this order in $\rho$, extension of the results to three dimensions presents no conceptual
difficulties. In particular, the mechanism of production for concentration fluctuations, and 
for correlation time and PDF corrections, is not expected to suffer modifications. Whether
a random velocity field model like the present one, could be appropriate to describe
transport by a three dimensional turbulence, laden with coherent structures and intermittency,
is a different matter.

The present extension to finite correlation times of the Kraichnan model is
perturbative in nature. Imposition of time statistics along Lagrangian 
trajectories had as consequence a non-Gaussian velocity field.
This resulted in a field theoretical perturbation theory, with expansion 
parameter $\rho^{-1}$, which is somewhat different from other field theories 
arising from closure analysis of the Navier Stokes equation. It would be interesting
to understand the relation with such theories, in particular with the 
quasi-Lagrangian approach described in \cite{belinicher97} and following papers based on this
work (see \cite{l'vov01} and references therein).

There are situations in which the higher orders in $\rho^{-1}$ become necessary. A relevant
example could be the derivation of a turbulent closure: in this case, extension of the theory 
to realistic values of $\rho$ could not be avoided.  Related to this issue, is the calculation 
of the anomalous scaling exponents for a passive scalar advected by a random velocity field
with finite correlation time. The analysis of
pair diffusion carried on in section III proceeded, at the end, as if the velocity field
had zero correlation time.
To lowest order in $\rho^{-1}$, the same zero-mode
structure of the Kraichnan model is therefore expected \cite{gawedzki95}. 
To proceed in a consistent
way, one should go to higher order, at the same time, in the passive tracer 
part of the problem and in the field theory for the velocity field.
Such issues, concerning the nature of the field theoretical perturbation expansion,
will be analyzed in a separate publication.

\vskip 5pt
\noindent{\bf Aknowledgements:} I wish to thank Paolo Muratore-Ginanneschi and Antti Kupiainen
for interesting and helpful conversation. Part of this research was carried on 
at the Mathematics Department of the University of Helsinki. I aknowledge support by 
a CNR Short Term Mobility fellowship.



\begin{thebibliography}{99}
\bibitem{note1} Going beyond lowest order would be necessary to evaluate the
nonlinear energy flux produced dynamically by Eqn. (2.7) and to see if there is an inverse
cascade, as in a real turbulent flow with similar characteristics.
\end{thebibliography}
\end{document}